\documentstyle[preprint,prd,eqsecnum,aps,epsf,axodraw]{revtex}

\tightenlines 
\begin{document}

    \newcommand{\DSC}{D\hspace{-0.25cm}\slash_{\bot}}
    \newcommand{\DSP}{D\hspace{-0.25cm}\slash_{\|}}   
    \newcommand{\DS}{D\hspace{-0.25cm}\slash}   
    \newcommand{\DC}{D_{\bot}}
    \newcommand{\DSCX}{D\hspace{-0.20cm}\slash_{\bot}}
    \newcommand{\DSPX}{D\hspace{-0.20cm}\slash_{\|}}  
    \newcommand{\DP}{D_{\|}}
    \newcommand{\QV}{Q_v^{+}}
    \newcommand{\QVB}{\bar{Q}_v^{+}}
    \newcommand{\QVP}{Q^{\prime +}_{v^{\prime}} }
    \newcommand{\QVBP}{\bar{Q}^{\prime +}_{v^{\prime}} }
    \newcommand{\QVHZ}{\hat{Q}^{+}_v}
    \newcommand{\QVHZB}{\bar{\hat{Q}}_v{\vspace{-0.3cm}\hspace{-0.2cm}{^{+}} } }    
    \newcommand{\QVPHZB}{\bar{\hat{Q}}_{v^{\prime}}{\vspace{-0.3cm}\hspace{-0.2cm}{^{\prime +}}} }    
    \newcommand{\QVPHFB}{\bar{\hat{Q}}_{v^{\prime}}{\vspace{-0.3cm}\hspace{-0.2cm}{^{\prime -}} } }    
    \newcommand{\QVPHB}{\bar{\hat{Q}}_{v^{\prime}}{\vspace{-0.3cm}\hspace{-0.2cm}{^{\prime}} }   }
    \newcommand{\QVHF}{\hat{Q}^{-}_v}
    \newcommand{\QVHFB}{\bar{\hat{Q}}_v{\vspace{-0.3cm}\hspace{-0.2cm}{^{-}} }}    
    \newcommand{\QVH}{\hat{Q}_v}
    \newcommand{\QVHB}{\bar{\hat{Q}}_v}
    \newcommand{\VS}{v\hspace{-0.2cm}\slash} 
    \newcommand{\MQ}{m_{Q}}
    \newcommand{\MQP}{m_{Q^{\prime}}}
    \newcommand{\QVHPMB}{\bar{\hat{Q}}_v{\vspace{-0.3cm}\hspace{-0.2cm}{^{\pm}} }}    
    \newcommand{\QVHMPB}{\bar{\hat{Q}}_v{\vspace{-0.3cm}\hspace{-0.2cm}{^{\mp}} }  }  
    \newcommand{\QVHPM}{\hat{Q}^{\pm}_v}
    \newcommand{\QVHMP}{\hat{Q}^{\mp}_v}

\draft
\title{ Semileptonic B Decays into Excited Charmed Mesons ($D_1$, $D^*_2$)
in HQEFT}
\author{ W.Y. Wang and Y.L. Wu }
\address{Institute of Theoretical Physics, Chinese Academy of Sciences, Beijing 100080, China } 
\maketitle

\begin{abstract} 
Exclusive semileptonic B decays into excited charmed mesons ($D_1$, $D^*_2$) are studied 
up to the order 
of $1/m_Q$ in the framework of the heavy quark effective field theory (HQEFT), 
which contains the contributions of both particles and antiparticles. 
Two wave functions $\eta^b_0$ and $\eta^c_0$, which characterize
the contributions from the kinematic operator at the order of $1/m_Q$,  are calculated by 
using QCD sum rule approach in HQEFT. Zero recoil values of 
other two wave functions $\kappa'_1$ and $\kappa'_2$ are extracted from the excited 
charmed-meson masses. Possible effects from the spin-dependent transition wave functions 
which arise from the magnetic operators at the order of $1/m_Q$ are analyzed.   
It is shown that the experimental measurements for the branching ratios 
of $B \rightarrow D_1\ l\nu$ and  $B \rightarrow D^*_2\ l\nu$  can be understood in 
the framework of HQEFT.

\end{abstract}


\newpage

\section{Introduction}\label{int}

Recently, studies on the semileptonic $B$ decays into excited charmed mesons become 
interesting in both experimental and theoretical sides. Experimentally, to precisely 
measure the branching ratios of the semileptonic $B$ meson into the groundstate 
charmed mesons, it also needs to measure precisely the branching ratios of 
the semileptonic $B$ decays into excited charmed mesons which is the main backgroud 
for the former decays. Theoretically, it provides additional modes for testing the 
validity of effective theories, in particular, how good of the spin-flavor symmetry.
 
The semileptonic $B$ decays into excited charmed mesons have been 
discussed in \cite{aklz,akl} based on the framework of the usual heavy quark effective 
theory (HQET).  Where the dependence of decay rates on wave functions 
was presented in a general form, the authors mainly deal with Isgur-Wise type function and 
the $1/m_Q$ order wave functions $\tau_1$ and $\tau_2$ which arise from the effective 
currents. The wave functions $\eta^Q_i$ arising from the chromomagnetic term 
in Lagrangian have been neglected, and those arising from the kinematic term in Lagrangian 
have been absorbed into the Isgur-Wise type function and not been considered separately. 
The detailed field theoretic calculations on these decays were presented in \cite{dh,dhl}, 
where the Isgur-Wise type function and $\tau_1$, $\tau_2$ have been evaluated 
by using the QCD sum rule approach. 

The purpose of this paper is to study the transitions between 
ground and excited state heavy-light mesons within the framework of 
heavy quark effective field theory (HQEFT)\cite{ylw} in which 
both quark and antiquark effective fields with keeping quark-antiquark coupled terms 
have been considered. In particular, 
we focus on the semileptonic B decays into the $j_l^P=\frac{3}{2}^+$ charmed meson 
doublet ($D_1$, $D^*_2$), where $j_l$ and $P$ are the spin and parity of the light degrees 
of freedom in the charmed mesons. It has been shown that the HQEFT can provide a consistent 
description on both exclusive \cite{wwy1,ww} and inclusive \cite{wwy2,wy} decays of heavy 
hadrons. It has been seen in these references that the HQEFT has many advantages 
with respect to the usual HQET. In the new framework, the values of $|V_{cb}|$ extracted 
from both exclusive and inclusive decays show a good agreement; the bottom hadron life time 
ratios can be well understood; $1/m_Q$ order corrections at zero recoil in both exclusive 
and inclusive decays automatically absent without imposing the equation of motion 
$iv\cdot D Q_v=0$ when the physical observables are presented in terms of heavy hadron masses; 
less wave functions are needed; and there are interesting relations among 
meson masses and transition wave functions at zero recoil. In the most recent paper \cite{ww}, 
the decay constants and binding energy of ground state heavy-light mesons as well as transition 
wave functions among them were studied consistently by using QCD sum rule approach in the 
framework of HQEFT. It was noticed that the HQEFT appears much more reliable than the usual 
HQET in describing the decays and transitions for the ground state mesons. 
Particularly, the $1/m_Q$ corrections to the heavy-light meson decay constants were found 
to be much smaller than the heavy quark mass, so that the scaling law of the decay constants 
is only slightly broken. This observation is unlike the usual HQET, which may lead 
to either the complete break down of the $1/m_Q$ expansion \cite{neu1076} or a large 
$1/m_Q$ correction \cite{ballnpb} in evaluating the decay constants. 
Two transition wave functions $\kappa_1(y)$ and $\kappa_2(y)$ at the $1/m_Q$ order have also 
been evaluated in \cite{ww}. Their zero recoil values agree with those 
extracted from the ground state meson masses. 

 Our present paper is organized as follows. 
In Sec.\ref{formulation}, the weak transition matrix elements relevant to $B\to (D_1, D^*_2)$ 
decays are studied up to the order of $1/m_Q$ and parameterized by independent universal 
wave functions. The form factors and decay rates are then given in terms of those 
wave functions. The relevant normalization of the excited mesons is also discussed. 
In Sec.\ref{sumrule}, using the appropriate interpolating currents for excited heavy 
mesons, we derive the sum rules for two of the important  wave functions concerned at $1/m_Q$ order, 
$\eta^b_0$ and $\eta^c_0$. In Sec.\ref{analysis}, we present our numerical results 
obtained from the sum rule approach. The $B\to (D_1,D^*_2)$ decay rates and branching ratios 
are discussed in detail. Finally, we come to our brief summary in Sec.\ref{sum}. 

 \section{Analytic formulae for $B\to (D_1,D^*_2)$ in HQEFT}\label{formulation}

The matrix elements relevant to the semileptonic decays $B\to (D_1,D^*_2)$ are the ones 
of vector and axial vector currents ($V^\mu=\bar c \gamma^\mu b$ and $A^\mu=\bar c 
\gamma^\mu \gamma^5 b$) between $B$ and the excited doublet ($D_1,D^*_2$). 
Usually, these matrix elements are parameterized as 
\begin{eqnarray}
\label{defformfactor}
 \langle D_1(v',\epsilon)|V^\mu|B(v) \rangle &=&\sqrt{m_{D_1} m_B} [f_{V_1}\epsilon^{*\mu}+
  ( f_{V_2}v^\mu+f_{V_3}v'^{\mu} ) (\epsilon^*\cdot v) ], \nonumber \\
 \langle D_1(v',\epsilon)|A^\mu|B(v) \rangle &=&i\sqrt{m_{D_1} m_B} f_A \epsilon^{\mu\alpha\beta\gamma} 
  \epsilon^*_\alpha v_\beta v'_\gamma, \nonumber \\
 \langle D^*_2(v',\epsilon)|A^\mu|B(v) \rangle &=&\sqrt{m_{D^*_2} m_B} [ k_{A_1} \epsilon^{*\mu\alpha} 
  v_\alpha+(k_{A_2}v^\mu+k_{A_3}v'^\mu ) \epsilon^*_{\alpha\beta} v^\alpha v^\beta ], \nonumber \\
 \langle D^*_2(v',\epsilon)|V^\mu|B(v) \rangle &=&i\sqrt{m_{D^*_2} m_B} k_V \epsilon^{\mu\alpha\beta\gamma}
  \epsilon^*_{\alpha\sigma} v^\sigma v_\beta v'_\gamma,
\end{eqnarray}
where the form factors $f_i$ and $k_i$ are dimensionless functions of $y=v\cdot v'$, 
and $\epsilon^{*\mu}$ ($\epsilon^{*\mu\alpha}$) is the polarization vector (tensor) 
of $D_1$ ($D^*_2$).

In the framework of HQEFT (its main formulation is presented in Appendix \ref{HQEFT}, 
we can in general introduce an effective heavy hadron state $|H_v \rangle $ for the witness of exhibiting 
manifestly the spin-flavor symmetry\cite{wwy1,ww,wwy2}. It is related to the hadron 
state $|H \rangle $ in the full theory via 
\begin{equation}
    \label{eq:statedef}    
     \frac{1}{\sqrt{m_{H^{\prime}}m_{H}}}  \langle H^{\prime}|\bar{Q}^{\prime} \Gamma Q \vert H \rangle = 
 \frac{1}{\sqrt{\bar{\Lambda}_{H^{\prime}} \bar{\Lambda}_H}}  \langle H^{\prime}_{v^{\prime}}\vert
          J_{eff} e^{i\int d^4x {\cal L}_{eff}} | H_v \rangle .
    \end{equation}  
with $|H_v \rangle $ being normalized as
   \begin{equation}
     \label{eq:statenor}
          \langle H_{v} \vert \QVB \gamma^{\mu} \QV \vert H_v \rangle  = 2\bar{\Lambda} v^{\mu} 
     \end{equation}
where 
\begin{equation}
\label{eq:bindrelat}
\bar{\Lambda} = \bar{\Lambda}_{H} - O(1/m_Q) = \lim_{m_{Q}\to \infty} \bar{\Lambda}_H 
\end{equation}
is taken to be heavy flavor independent, and it mainly reflects the effects of the light degrees of freedom 
in the heavy hadron characterizes the off-mass shellness of the heavy quark within the heavy hadron. 
Once $\bar{\Lambda}$ is chosen to be the flavor independent binding energy of the hadron $|H \rangle $, 
one yields
    \begin{equation}
    \label{eq:binding}
       \bar{\Lambda}_H \equiv m_H-m_Q.
    \end{equation}
which is the total binding energy of hadron. 

The hadronic matrix elements can be expanded according to the order 
of $1/m_Q$. By including the $1/m_Q$ order corrections which can arise from both the current expansion and 
the insertion of the effective Lagrangian, one gets \cite{wwy1,ww}
 \begin{eqnarray}
 \label{eq:matrixexp}
 &&\sqrt{\frac{\bar{\Lambda}_M}{m_M}}  \langle 0|\bar{q}\Gamma Q|M \rangle  \to 
   \langle 0|\bar{q} \Gamma \QV|M_v \rangle -\frac{1}{2m_Q}  \langle 0|\bar{q}\Gamma\frac{1}{i \DSP}(i\DSC)^2\QV |M_v \rangle 
   +O(1/m^2_Q), \nonumber \\
&& \sqrt{\frac{\bar{\Lambda}_{M'} \bar{\Lambda}_{M} }{m_{M'} m_M}}  \langle M'|\bar{Q}' \Gamma Q|M \rangle  \to 
  \langle M'_{v'}|\QVBP \Gamma \QV|M_v \rangle -\frac{1}{2m_Q}  \langle M'_{v'}|\QVBP\Gamma\frac{1}{i \DSP}(i\DSC)^2\QV |M_v \rangle 
\nonumber\\
&&\hspace{2cm}  -\frac{1}{2m_{Q'}}  \langle M'_{v'}|\QVBP (-i\stackrel{\hspace{-0.1cm}\leftarrow}{\DSC})^2
   \frac{1}{-i\stackrel{\hspace{-0.1cm}\leftarrow}{\DSP}}
   \Gamma \QV|M_v \rangle  +O(1/m^2_Q). 
 \end{eqnarray}

The form factors $f_i$ and $k_i$ can be parameterized by a set of universal wave 
functions. It is simplest to do this by using the trace formulism \cite{aff,akl}.
The spin wave functions for the $j_l^P=\frac{1}{2}^-$ ground state mesons $B$, $B^*$ 
and $j_l^P=\frac{3}{2}^+$ charmed mesons $D_1$, $D^*_2$ are 
   \begin{eqnarray}
     \label{eq:spinwave1}
       {\cal M}_v=\sqrt{\bar{\Lambda}}P_{+}
         \left\{
           \begin{array}{cl}
              -\gamma^{5}, & \mbox{for pseudoscalar meson} \; \; \\
              \epsilon\hspace{-0.15cm}\slash, & \mbox{for vector meson} \; \; 
           \end{array}
         \right.
     \end{eqnarray}
\begin{eqnarray}
  \label{eq:spinwave2}
    {\cal F}^\mu_v=\sqrt{\bar{\Lambda}'}P_{+}
      \left\{
        \begin{array}{cl}
          -\sqrt{\frac{3}{2}} \gamma^5 \epsilon^\nu [g^\mu_\nu-\frac{1}{3}
            \gamma_\nu (\gamma^\mu-v^\mu) ], & \mbox{for $D_1$} \;\;  \\
          \epsilon^{\mu\nu} \gamma_\nu, & \mbox{for $D^*_2$} \; \; 
        \end{array}
      \right.
  \end{eqnarray}
The matrices ${\cal{M}}_v$ and ${\cal{F}}^\mu_v$ 
satisfy the properties $\VS {\cal{M}}_v={\cal{M}}_v=-{\cal{M}}_v \VS$, 
$\VS {\cal{F}}^\mu_v={\cal{F}}^\mu_v=-{\cal{F}}^\mu_v \VS$ and 
${\cal{F}}^\mu_v \gamma_\mu={\cal{F}}^\mu_v v_\mu=0$.

With respect to eq.(\ref{eq:statenor}) and the normalization of the $\frac{3}{2}^+$ 
excited states, we parameterize the matrix elements in eq.(\ref{eq:matrixexp}) as follows 
\begin{eqnarray}
 \langle H_v|\QVB \gamma^\mu \QV|H_v \rangle &=& \xi'(y) Tr[\bar{\cal{F}}^\sigma_v \gamma^\mu 
    {\cal{F}}_{v \sigma} ], \nonumber\\
 \langle H_v|\QVB \gamma^\mu \frac{P_+}{i v\cdot D} \DC^2 \QV|H_v \rangle &=& -\kappa'_1(y) 
    \frac{1}{\bar{\Lambda}'}
    Tr[\bar{\cal{F}}^\sigma_v \gamma^\mu {\cal{F}}_{v \sigma} ] ,\nonumber\\
 \langle H_v|\QVB \gamma^\mu \frac{P_+}{i v\cdot D} \frac{i}{2} \sigma_{\alpha \beta} F^{\alpha\beta}
   \QV|H_v \rangle &=& \frac{1}{\bar{\Lambda}'}
  Tr[ A^{\sigma\sigma'\alpha\beta}(v,v') \bar{{\cal{F}}}_{v \sigma'} \gamma^\mu P_+
    \frac{i}{2} \sigma_{\alpha\beta} {\cal{F}}_{v \sigma} ] 
\end{eqnarray}
with 
\begin{eqnarray}
A^{\sigma\sigma'\alpha\beta}(v,v')&=&a_1 (g^{\sigma\alpha}g^{\sigma'\beta}
  -g^{\sigma\beta}g^{\sigma'\alpha})+i a_2 g^{\sigma\sigma'} \sigma^{\alpha\beta} 
  +a_3 g^{\sigma\sigma'} (v^\alpha \gamma^\beta -v^\beta \gamma^\alpha) \nonumber \\
  &+& ia_4 (g^{\sigma\alpha} \sigma^{\sigma'\beta}-g^{\sigma\beta} \sigma^{\sigma'\alpha}) 
  +a_5 (g^{\sigma\alpha} v^\beta - g^{\sigma\beta} v^\alpha )\gamma^{\sigma'}
  +ia_6 (g^{\sigma'\alpha}\sigma^{\sigma\beta}-g^{\sigma'\beta}\sigma^{\sigma\alpha} )\nonumber\\
  &+&a_7 (g^{\sigma'\alpha} v^\beta -g^{\sigma'\beta} v^\alpha)\gamma^\sigma 
  +a_8 \sigma^{\sigma'\sigma} \sigma^{\alpha\beta}
  +ia_9 \sigma^{\sigma'\sigma} (\gamma^\alpha v^\beta-\gamma^\beta v^\alpha).
\end{eqnarray}
Finishing the trace calculations, we get from (\ref{eq:statedef}) and (\ref{eq:statenor})
\begin{eqnarray}
\label{eq:mQnor}
2m_{D_1} v^\mu&=&\frac{m_{D_1}}{\bar{\Lambda}_{D_1}} \{-2\bar{\Lambda}' \xi'+\frac{2}{m_c}
  (\kappa'_1(1)+5\kappa'_2(1)) \} (\epsilon^* \cdot \epsilon) v^\mu  , \nonumber\\
2m_{D^*_2} v^\mu&=&\frac{m_{D^*_2}}{\bar{\Lambda}_{D^*_2}} \{2\bar{\Lambda}' \xi'-\frac{2}{m_c}
  (\kappa'_1(1)-3\kappa'_2(1))  \} \epsilon^{*\sigma\alpha} \epsilon_{\sigma\alpha} v^\mu,
\end{eqnarray}
where 
\begin{equation}
\kappa'_2=-\frac{1}{3} (a_1-a_2-a_4-a_6-a_8)
\end{equation}
Eq.(\ref{eq:mQnor}) yields
\begin{eqnarray}
\label{lamb1rel}
\bar{\Lambda}_{D_1} &=& \bar{\Lambda}' -\frac{1}{m_c} 
   (\kappa'_1(1)+5\kappa'_2(1)), \nonumber\\
\bar{\Lambda}_{D^*_2} &=& \bar{\Lambda}' -\frac{1}{m_c} 
   (\kappa'_1(1)-3\kappa'_2(1)) .
\end{eqnarray}
In deriving these formulae we have used the normalization of the leading order wave 
function $\xi'(1)$:
\begin{equation}
 \xi'(1)=1 ,
\end{equation}
which is a direct result of eqs.(\ref{eq:bindrelat}) and (\ref{eq:mQnor}). 

Eq.(\ref{lamb1rel}) is similar to those relations for ground state mesons \cite{wwy1,ww}: 
\begin{eqnarray}
\label{lambrel}
\bar{\Lambda}_{D} &=& \bar{\Lambda} -\frac{1}{m_c} 
   (\kappa_1(1)+3\kappa_2(1)) ,\nonumber\\
\bar{\Lambda}_{D^*} &=& \bar{\Lambda} -\frac{1}{m_c} 
   (\kappa_1(1)-\kappa_2(1)) .
\end{eqnarray}

For $B\to (D_1,D^*_2)$ decays, we parameterize the relevant matrix elements as follows 
\begin{eqnarray}
 \langle H_{v'}|\QVBP \Gamma \QV|H_v \rangle &=&\tau(y) Tr[v_\sigma \bar{{\cal{F}}}^\sigma_{v'} \Gamma 
   {\cal{M}}_v ]  ,\nonumber \\
 \langle H_{v'}|\QVBP \Gamma \frac{P_+}{i v\cdot D } \DC^2 \QV|H_v \rangle &=&-\eta^b_0(y) \frac{1}{\bar{\Lambda}} 
   Tr[v_\sigma \bar{{\cal{F}}}^\sigma_{v'} \Gamma {\cal{M}}_v ] , \nonumber \\   
 \langle H_{v'}|\QVBP {\stackrel{\hspace{-0.1cm}\leftarrow}\DC}^2 
   \frac{P'_+}{-i v' \cdot \stackrel{\hspace{-0.1cm}\leftarrow}D }
\Gamma \QV|H_v \rangle 
  &=&-\eta^c_0(y) \frac{1}{\bar{\Lambda}'} 
   Tr[v_\sigma \bar{{\cal{F}}}^\sigma_{v'} \Gamma {\cal{M}}_v ] , \nonumber \\   
 \langle H_{v'}|\QVBP \Gamma \frac{P_+}{i v\cdot D } \frac{i}{2} \sigma_{\alpha\beta} F^{\alpha\beta} 
   \QV|H_v \rangle &=& -\frac{1}{\bar{\Lambda}} Tr[ R^b_{\sigma\alpha\beta} (v,v')
   \bar{\cal{F}}_{v'}^\sigma \Gamma P_+ i\sigma_{\alpha\beta} {\cal{M}}_v ] ,\nonumber \\   
 \langle H_{v'}|\QVBP \frac{i}{2}\sigma_{\alpha\beta} F^{\alpha\beta} 
   \frac{P'_+}{-i v'\cdot \stackrel{\hspace{-0.1cm}\leftarrow} D}
\Gamma \QV|H_v \rangle 
    &=&-\frac{1}{\bar{\Lambda}'} Tr[ R^c_{\sigma\alpha\beta} (v,v')
   \bar{\cal{F}}_{v'}^\sigma i\sigma^{\alpha\beta} P'_+ \Gamma {\cal{M}}_v ] ,
\end{eqnarray}
where $R^b_{\sigma\alpha\beta}$ and $R^c_{\sigma\alpha\beta}$ can generally be written 
as 
\begin{eqnarray}
R^b_{\sigma\alpha\beta}(v,v')&=&\eta^b_1 v_\sigma \gamma_\alpha \gamma_\beta 
  +\eta^b_2 v_\sigma v'_\alpha \gamma_\beta+\eta^b_3 g_{\sigma\alpha} v'_{\beta} ,\nonumber \\
R^c_{\sigma\alpha\beta}(v,v')&=&\eta^c_1 v_\sigma \gamma_\alpha \gamma_\beta 
  +\eta^c_2 v_\sigma v_\alpha \gamma_\beta+\eta^c_3 g_{\sigma\alpha} v_{\beta} .
\end{eqnarray}
$\tau$, $\eta^b_i$ and $\eta^c_i$ are Lorentz scalar wave functions of $y$. 

Now finishing the trace calculations, and taking into account 
eqs.(\ref{eq:statedef}), (\ref{defformfactor}), (\ref{eq:matrixexp}), 
(\ref{lamb1rel}) and (\ref{lambrel}), one finds that $f_i$ and $k_i$ are related to 
the transition wave functions as follows 
\begin{eqnarray}
\label{factorfunction}
\sqrt{6} f_{V_1}&=&(1-y^2) \{ [\tilde{\tau}+\frac{\tau}{2m_b\bar{\Lambda} }
  (\kappa_1(1)+3\kappa_2(1))+\frac{\tau}{2m_c \bar{\Lambda}'} 
  (\kappa'_1(1)+5\kappa'_2(1)) ] \nonumber\\
  &+&\frac{1}{m_b\bar{\Lambda}} \eta^b 
  +\frac{1}{m_c \bar{\Lambda}'} [\eta^c_1+\frac{3}{2} \eta^c_3] \}, \nonumber \\
\sqrt{6} f_{V_2}&=&-3 [\tilde{\tau}+\frac{\tau}{2m_b\bar{\Lambda} }
  (\kappa_1(1)+3\kappa_2(1))+\frac{\tau}{2m_c \bar{\Lambda}'} 
  (\kappa'_1(1)+5\kappa'_2(1)) ] \nonumber\\
  &-&\frac{3}{m_b\bar{\Lambda}} \eta^b 
  -\frac{5}{m_c \bar{\Lambda}'} [-\eta^c_1+\frac{1}{2} \eta^c_3] , \nonumber \\
\sqrt{6} f_{V_3}&=&(y-2) [\tilde{\tau}+\frac{\tau}{2m_b\bar{\Lambda} }
  (\kappa_1(1)+3\kappa_2(1))+\frac{\tau}{2m_c \bar{\Lambda}'} 
  (\kappa'_1(1)+5\kappa'_2(1)) ] \nonumber \\
  &+&\frac{1}{m_b\bar{\Lambda}} (y-2) \eta^b 
  +\frac{1}{m_c \bar{\Lambda}'} [\eta^c_1 (6+y)-2 \eta^c_2 (1-y) 
  -\eta^c_3 (1-\frac{3}{2} y)   ] , \nonumber \\
\sqrt{6} f_{A}&=&-(1+y) \{ [\tilde{\tau}+\frac{\tau}{2m_b\bar{\Lambda} }
  (\kappa_1(1)+3\kappa_2(1))+\frac{\tau}{2m_c \bar{\Lambda}'} 
  (\kappa'_1(1)+5\kappa'_2(1)) ]  \nonumber \\
  &+&\frac{1}{m_b\bar{\Lambda}} \eta^b 
  +\frac{1}{m_c \bar{\Lambda}'} [\eta^c_1+\frac{3}{2} \eta^c_3] \}, \nonumber \\
k_{A_1}&=&-(1+y) \{ [\tilde{\tau}+\frac{\tau}{2m_b\bar{\Lambda} }
  (\kappa_1(1)+3\kappa_2(1))+\frac{\tau}{2m_c \bar{\Lambda}'} 
  (\kappa'_1(1)-3\kappa'_2(1)) ] \nonumber \\
  &+&\frac{1}{m_b\bar{\Lambda}} \eta^b 
  +\frac{1}{m_c \bar{\Lambda}'} [\eta^c_1-\frac{1}{2} \eta^c_3] \}, \nonumber \\
k_{A_2}&=&\frac{1}{m_c \bar{\Lambda}'} \eta^c_2, \nonumber \\
k_{A_3}&=& [\tilde{\tau}+\frac{\tau}{2m_b\bar{\Lambda} }
  (\kappa_1(1)+3\kappa_2(1))+\frac{\tau}{2m_c \bar{\Lambda}'} 
  (\kappa'_1(1)-3\kappa'_2(1)) ] \nonumber \\
  &+&\frac{1}{m_b\bar{\Lambda}} \eta^b 
  +\frac{1}{m_c \bar{\Lambda}'} [\eta^c_1-\eta^c_2-\frac{1}{2} \eta^c_3] , \nonumber \\
k_{V}&=& -[\tilde{\tau}+\frac{\tau}{2m_b\bar{\Lambda} }
  (\kappa_1(1)+3\kappa_2(1))+\frac{\tau}{2m_c \bar{\Lambda}'} 
  (\kappa'_1(1)-3\kappa'_2(1)) ]  \nonumber \\
  &-&\frac{1}{m_b\bar{\Lambda}} \eta^b 
  -\frac{1}{m_c \bar{\Lambda}'} [\eta^c_1-\frac{1}{2} \eta^c_3] ,
\end{eqnarray}
where 
\begin{eqnarray}
 \tilde{\tau}&=&\tau-\frac{\eta^b_0}{2m_b \bar{\Lambda}}-\frac{\eta^c_0}{2m_c 
    \bar{\Lambda}'}, \nonumber \\
 \eta^b&=&-3\eta^b_1-(1-y) \eta^b_2-\frac{1}{2}\eta^b_3.
\end{eqnarray}
 
 \section{QCD sum rule Evaluation for $\tau$, $\eta^b_0$ and $\eta^c_0$}\label{sumrule}

In order to calculate the wave functions $\tau$ and $\eta^b_0$, $\eta^c_0$, one may 
study the analytic properties of the three-point correlation functions
\begin{eqnarray}
\label{correlator1}
&&i^2 \int d^4x d^4z e^{i (p'\cdot x-p\cdot z)}  \langle 0|T\{ J^{\nu}_{1,+,3/2} (x), 
  (\bar{Q}\Gamma Q)(0), J^{\dagger}_{0,-,1/2}(z) \}|0 \rangle  = \Xi^\tau (\omega,\omega',y) L^{\mu\nu}_{V,A}, \\
\label{correlator2}
&&i^2 \int d^4x d^4z e^{i (p'\cdot x-p\cdot z)}  \langle 0|T\{ J^{\alpha\beta}_{2,+,3/2}(x), 
  (\bar{Q}\Gamma Q)(0), J^{\dagger}_{0,-,1/2}(z) \}|0 \rangle  = \Xi^\tau (\omega,\omega',y) 
  L^{\mu\alpha\beta}_{V,A}, \\
  \label{correlator3}
&&i^2 \int d^4x d^4z e^{i (p'\cdot x-p\cdot z)}  \langle 0|T\{ J^{\nu}_{1,+,3/2}(x), 
  (\bar{Q}\Gamma \frac{P_+}{iv\cdot D} \DC^2 Q)(0), J^{\dagger}_{0,-,1/2}(z) \}|0 \rangle  \nonumber\\
 &&\hspace{4cm} = \Xi^{\eta^b_0} (\omega,\omega',y) L^{\mu\nu}_{V,A}, \\
  \label{correlator4}
&&i^2 \int d^4x d^4z e^{i (p'\cdot x-p\cdot z)}  \langle 0|T\{ J^{\alpha\beta}_{2,+,3/2}(x), 
  (\bar{Q}\Gamma \frac{P_{+}}{iv\cdot D} \DC^2 Q)(0), J^{\dagger}_{0,-,1/2}(z) \}|0 \rangle  \nonumber\\
 &&\hspace{4cm}  = \Xi^{\eta^b_0} (\omega,\omega',y) L^{\mu\alpha\beta}_{V,A}, \\
  \label{correlator5}
&&i^2 \int d^4x d^4z e^{i (p'\cdot x-p\cdot z)}  \langle 0|T\{ J^{\nu}_{1,+,3/2}(x), 
  (\bar{Q} \stackrel{\hspace{-0.2cm}\leftarrow}{\DC^2} \frac{P'_{+}}
  {-iv'\cdot \stackrel{\hspace{-0.2cm}\leftarrow} D }\Gamma Q)(0),
  J^{\dagger}_{0,-,1/2}(z) \}|0 \rangle  \nonumber \\
 &&\hspace{4cm}   = \Xi^{\eta^c_0}(\omega,\omega',y) L^{\mu\nu}_{V,A}, \\
  \label{correlator6}
&&i^2 \int d^4x d^4z e^{i (p'\cdot x-p\cdot z)}  \langle 0|T\{ J^{\alpha\beta}_{2,+,3/2}(x), 
  (\bar{Q} \stackrel{\hspace{-0.2cm}\leftarrow}{\DC^2}
  \frac{P'_+}{-iv'\cdot \stackrel{\hspace{-0.2cm}\leftarrow} D } \Gamma Q)(0),
  J^{\dagger}_{0,-,1/2}(z) \}|0 \rangle  \nonumber\\
 &&\hspace{4cm}  = \Xi^{\eta^c_0}(\omega,\omega',y) L^{\mu\alpha\beta}_{V,A}
\end{eqnarray}
with $\Gamma$ being $\gamma^\mu$ and $\gamma^\mu \gamma^5$ for vector and axial vector 
heavy quark currents separately. $L^{\mu\nu(\mu\alpha\beta)}_{V,A}$ are Lorentz structures 
associated with the vector and axial vector currents (see Appendix \ref{Los}).

$J^{\nu (\alpha\beta)}_{j,P,j_l}$ are the proper interpolating currents for the heavy-light 
mesons. Here
\begin{eqnarray}
J^{\dagger\alpha}_{0,-,1/2}&=&\sqrt{\frac{1}{2}} \bar{Q} \gamma^5 q \hspace{3cm} 
\mbox{for pseudoscalar meson}, \nonumber \\
J^{\dagger\alpha}_{1,-,1/2}&=&\sqrt{\frac{1}{2}}\bar{Q} \gamma^\alpha_\bot q \hspace{3cm} 
\mbox{for vector meson},
\end{eqnarray}
for $\frac{1}{2}^-$ ground state doublet, and 
\begin{eqnarray}
J^{\dagger \alpha}_{1,+,3/2}=\sqrt{\frac{3}{4}}\QVB \gamma^5 (-i)(D^{\alpha}_\bot 
  -\frac{1}{3} \gamma^{\alpha}_\bot \DS_\bot )q \hspace{3cm} \mbox{for $D_1$}, \nonumber \\
J^{\dagger \alpha\beta}_{2,+,3/2}=\sqrt{\frac{1}{2}}\QVB \frac{(-i)}{2} (\gamma^\alpha_\bot 
  D^\beta_\bot + \gamma^\beta_\bot D^\alpha_\bot -\frac{2}{3} g^{\alpha\beta}_\bot \DS_\bot )q \hspace{3cm}
  \mbox{for $D^*_2$}.
\end{eqnarray}
 for $\frac{3}{2}^+$ doublet.

Generally, the proper current $J_{j,P,j_l}$ for the state with quantum numbers 
$j$, $P$, $j_l$ have been investigated in \cite{dhh,dh}. These currents were proved 
to satisfy the following conditions
\begin{eqnarray}
\label{currentfea}
 \langle 0|J^{\alpha_1\cdots\alpha_j}_{j,P,j_l}(0)|j',P',j'_l \rangle &=&if_{Pj_l}\delta_{jj'}
  \delta_{PP'} \delta_{j_lj'_l} \eta^{\alpha_1\cdots\alpha_j}, \nonumber \\
i \langle 0|T(J^{\alpha_1\cdots\alpha_j}_{j,P,j_l}(x) J^{\dagger \beta_1\cdots\beta_{j'}}_{j',P',j'_l}(0))
  |0 \rangle &=&\delta_{jj'} \delta_{PP'} \delta_{j_lj'_l} (-1)^j {\cal{S}} g^{\alpha_1\beta_1}_\bot
  \cdots g^{\alpha_j \beta_j}_\bot \nonumber \\
  \times \int dt \delta(x-vt) \Pi_{P,j_l}(x)
\end{eqnarray}
in the limit $m_Q\to \infty$. $\eta^{\alpha_1\cdots\alpha_j}$ is the polarization 
tensor for the spin $j$ state, 
$g^{\alpha\beta}_\bot=g^{\alpha\beta}-v'^{\alpha}v'^{\beta}$ is the transverse metric tensor, 
and $\gamma^{\alpha}_\bot=\gamma^\alpha-v^\alpha (v\cdot \gamma)$. 
$\cal{S}$ denotes symmetrizing the indices and subtracting the trace terms separately 
in the sets ($\alpha_1 \cdots \alpha_j$) and ($\beta_1 \cdots \beta_j$), $f_{P,j_l}$ 
and $\Pi_{P,j_l}$ are a constant and a function of $x$ respectively, they depend only 
on $P$ and $j_l$. 

Eqs.(\ref{currentfea}) implies that the sum rules in HQET (or HQEFT) for decay 
amplitudes derived from correlators containing such currents receive contributions 
only from one of the two states with the same spin-parity ($j$, $P$) in the $m_Q\to \infty$ 
limit. And starting from the leading order, the $1/m_Q$ corrections to the decay amplitudes 
can then be calculated unambiguously order by order. 

It should be noticed that the HQEFT differs from the HQET only in the $1/m_Q$ corrections. 
Therefore $f_{P,j_l}$ are the same in the two frameworks. For the ground state 
$\frac{1}{2}^-$ mesons, the sum rule for $f_{-,1/2}$ is also known \cite{neu1076,neurep} 
in the HQET. It was also checked again in the HQEFT (where $F=\sqrt{2} f_{-,1/2}$)\cite{ww} . 
Our result is 
\begin{eqnarray}
\label{sr1}
f^2_{-,1/2} e^{-2\bar{\Lambda}/T}&=&\frac{3}{16\pi^2} \int^{\omega_c}_{0} d\omega 
  e^{-\omega/T} \omega^2 -\frac{ \langle \bar{q}q \rangle }{2} (1+\frac{4\alpha_s}{3\pi}) \nonumber \\
 && -i\frac{ \langle \bar{q}\sigma_{\alpha\beta}F^{\alpha\beta} q \rangle }{8T^2}
  ( 1+\frac{4 \alpha_s}{\pi} ) -\frac{ \langle \alpha_sFF \rangle }{48 \pi T}={\cal SR}_{-,1/2}.
\end{eqnarray}
For the $\frac{3}{2}^+$ doublet, the sum rule for $f_{+,3/2}$ is found to be \cite{dhh}
\begin{eqnarray}
\label{sr2}
f^2_{+,3/2} e^{-2\bar{\Lambda}'/T}=\frac{1}{64\pi^2} \int^{\omega^{**}_c}_{0} d\omega 
  e^{-\omega/T} \omega^4 +i\frac{ \langle \bar{q}\sigma_{\alpha\beta}F^{\alpha\beta} q \rangle }{12} 
  -\frac{ \langle \alpha_sFF \rangle  T}{32 \pi}={\cal SR}_{+,3/2}.
\end{eqnarray}

The leading order wave function $\tau$ for $B\to (D_1,D^*_2)$ decays is evaluated in 
\cite{dh} through studying the three point correlation functions (\ref{correlator1}), 
and (\ref{correlator2}). The Borel transformed sum rule for it reads 
\begin{eqnarray}
\label{srtau}
f_{-,1/2} f_{+,3/2} \tau e^{-(\bar{\Lambda}+\bar{\Lambda}')/T}&=&\frac{1}{2\pi^2} 
  \frac{1}{(y+1)^3} \int^{\omega^{*}_c}_{0} d\omega 
  e^{-\omega/T} \omega^3 \nonumber\\
  &+&i\frac{ \langle \bar{q}\sigma_{\alpha\beta}F^{\alpha\beta} q \rangle }{12T}
  -\frac{ \langle \alpha_sFF \rangle }{96 \pi} \frac{y+5}{(y+1)^2}={\cal SR}_{\tau}.
\end{eqnarray}

The total external momenta in (\ref{correlator1})-(\ref{correlator6}) are 
$p=m_Qv+k$ and $p'=m_{Q'}v'+k'$ with $k$ and $k'$ being the residual momenta of the 
heavy quarks. 
$\Xi^{\eta^b_0}(\omega,\omega',y)$ and $\Xi^{\eta^c_0}(\omega,\omega',y)$ are analytic functions of 
$\omega=2v\cdot k+O(1/m_Q)$ and $\omega'=2v'\cdot k'+O(1/m_Q')$ with discontinuities 
for their positive values. Saturating the correlators in eqs.(\ref{correlator3})-(\ref{correlator6}) 
with physical intermediate states 
in HQEFT, phenomenologically we represent them as follows
\begin{eqnarray}
\label{phenb}
\Xi^{\eta^b_0}_{phen} L^{\mu\alpha(\mu\alpha\beta)}_{V,A}&=&\sum_{M',M}
   \frac{ \langle 0|J^{\alpha(\alpha\beta)}_{j,+,3/2}|M' \rangle 
    \langle M'|\bar{Q}\Gamma\frac{P_{+}}{iv\cdot D}\DC^2 Q|M \rangle  \langle M|J^{\dagger}_{0,-,1/2} |0 \rangle  } 
   {(2\bar{\Lambda}-\omega-i\epsilon)(2\bar{\Lambda}'-\omega'-i\epsilon)}\nonumber\\
   &+&\int_D d\nu d\nu' \frac{\rho_{phys}(\nu,\nu')}{(\nu-\omega-i\epsilon)(\nu'-\omega'-i\epsilon)}
   L^{\mu\alpha(\mu\alpha\beta)}_{V,A}
   +subtractions ,\\
\label{phenc}
\Xi^{\eta^c_0}_{phen}L^{\mu\alpha(\mu\alpha\beta)}_{V,A}&=&\sum_{M',M}
  \frac{ \langle 0|J^{\alpha(\alpha\beta)}_{j,+,3/2}|M' \rangle 
   \langle M'|(\bar{Q} \stackrel{\hspace{-0.2cm}\leftarrow}{\DC^2} \frac{P'_+}
  {-iv'\cdot \stackrel{\hspace{-0.2cm}\leftarrow} D}\Gamma Q)_{(0)}|M \rangle   \langle M|J^{\dagger}_{0,-,1/2} |0 \rangle  }
 {(2\bar{\Lambda}-\omega-i\epsilon)(2\bar{\Lambda}'-\omega'-i\epsilon)} \nonumber\\
  &+&\int_D d\nu d\nu' \frac{\rho_{phys}(\nu,\nu')}{(\nu-\omega-i\epsilon)(\nu'-\omega'-i\epsilon)}
  L^{\mu\alpha(\mu\alpha\beta)}_{V,A}
  +subtractions
\end{eqnarray}
with the first term in each equation being a double-pole contribution, and the 
second representing the higher resonance contributions in the form of a double dispersion 
integral over physical intermediate states in the proper integration domain $D$. 

The matrix elements in (\ref{phenb}) and (\ref{phenc}) can then be transformed into series 
of the matrix elements in the HQEFT in powers of $1/m_Q$ through (\ref{eq:matrixexp}). 
With (\ref{currentfea}), the first terms in (\ref{phenb}) and (\ref{phenc}) become in the 
limit $m_Q\to \infty$
\begin{eqnarray}
\label{poleb}
\Xi^{\eta^b_0}_{pole} L^{\mu\alpha(\mu\alpha\beta)}_{V,A}&=&
  \frac{f_{+,3/2}f_{-,1/2}}{(2\bar{\Lambda}-\omega-i\epsilon)(2\bar{\Lambda}'-\omega'-i\epsilon)}
  \frac{-\eta^b_0}{\bar{\Lambda}} L^{\mu\alpha(\mu\alpha\beta)}_{V,A}  \\
\label{polec}
\Xi^{\eta^c_0}_{pole} L^{\mu\alpha(\mu\alpha\beta)}_{V,A}&=&
  \frac{f_{+,3/2}f_{-,1/2}}{(2\bar{\Lambda}-\omega-i\epsilon)(2\bar{\Lambda}'-\omega'-i\epsilon)}
  \frac{-\eta^c_0}{\bar{\Lambda}'} L^{\mu\alpha(\mu\alpha\beta)}_{V,A}
\end{eqnarray}

On the other hand, the correlation functions may be written as 
\begin{eqnarray}
\label{theo}
\Xi^{\eta^{b(c)}_0}_{theo}=\int^{\infty}_{-\infty} d\nu d\nu' \frac{\rho^{b(c)}_{pert}} 
   {(\nu-\omega-i\epsilon)(\nu'-\omega'-i\epsilon)} +\Xi_{NP} +subtractions
\end{eqnarray}
with the first term being perturbative contributions and the second non-perturbative 
ones. These can be calculated order by order in the framework of HQEFT by using the 
perturbation theory and the operator product expansion (OPE) as well. The Lorentz 
structure $L^{\mu\alpha(\mu\alpha\beta)}_{V,A}$ is extracted out and not presented in 
(\ref{theo}). 

Assuming the quark-hadron duality, the sum rules for $\eta^b_0$ and $\eta^c_0$ read 
\begin{eqnarray}
\label{poletheo}
\Xi^{\eta^{b(c)}_0}_{pole}=\int^{\nu_1}_{0} d\nu \int^{\nu_2}_{0} d\nu' 
   \frac{\rho^{b(c)}_{pert}}{(\nu-\omega-i\epsilon)(\nu'-\omega'-i\epsilon)} 
   +\Xi_{NP} + subtractions.
\end{eqnarray}

The next step of sum rule method is to perform the Borel operator 
\begin{eqnarray}
 \label{eq:defBT}
   \hat{B}^{(\omega)}_{T}\equiv T  \lim_{n \to \infty,-\omega \to \infty} 
   \frac{ \omega^n}{\Gamma(n)} (-\frac{d}{d\omega})^n 
   \;\;\; \mbox{with} \;\; T=\frac{-\omega}{n} \;\; \mbox{fixed} 
 \end{eqnarray}
 to both sides of sum rules. Since there are two momentum variables $\omega$ and 
 $\omega'$ for the correlators (\ref{correlator3})-(\ref{correlator6}), 
a double Borel transformation $\hat{B}^{(\omega')}_{t'} \hat{B}^{(\omega)}_{t}$ 
should be performed to them. This has the effect to suppress 
higher resonance contributions on one hand, and to enhance the importance of low 
dimension condensates on the other, and thirdly, it also eliminates the subtraction 
terms. 

In QCD sum rule analysis for B semileptonic decays into ground state charmed 
mesons, it was argued \cite{neurep,bs,neubprd46} that the 
hadronic and perturbative spectral densities can not be locally dual to each other. 
But if one integrates the spectral densities over the "off-diagonal" variable 
$\nu_{-}=\frac{\nu-\nu'}{2}$, keeping the "diagonal" variable $\nu_{+}=\frac{\nu+\nu'}{2}$ 
fixed, the quark-hadron duality is restored in $\nu_{+}$ for the integrated 
spectral densities. This method was also used in \cite{ww} to calculate the transitions 
between ground state mesons in the HQEFT. 

In the present case, the initial and final states belong to different doublets and 
are asymmetric. If one uses an asymmetric triangle in the perturbative integral, however, 
the resulting wave functions or their derivatives will unfortunately be divergent 
at $y=1$ \cite{dh}.
For these reasons, here we shall follow the method used in \cite{ww,dh}, namely taking $t'=t=2T$ and
integrating first the spectral density over $\nu_{-}$ in the region 
$-\nu_{+} \langle \nu_{-} \langle \nu_{+}$, we then obtain from the quark-hadron duality the following form
  \begin{eqnarray}
  \label{eq:IWphentheo}
   \tilde{\Xi}_{pole}=2\int^{\omega^{*(1)}_c}_{0} d\nu_{+} e^{-\nu_{+}/T} 
      \tilde{\rho}_{pert}(\nu_{+})+\tilde{\Xi}_{NP},
  \end{eqnarray}
 where $\tilde{\Xi}$ denotes the result obtained by applying double Borel operators to $\Xi$, 
and 
  \begin{eqnarray}
    \tilde{\rho}_{pert}(\nu_{+})=\int^{\nu_{+}}_{-\nu_{+}} d\nu_{-} \rho_{pert}(\nu_{+},\nu_{-}).
  \end{eqnarray}

In the present calculations, we will neglect the light quark mass and higher radiative 
corrections. For non-perturbatve terms, we consider only the contributions of the 
quark condensate, the gluon condensate and the mixed quark-gluon condensate. 
The relevant Feynman diagrams are presented in Fig.1.
From the sum rules, we arrive at the following results
\begin{eqnarray}
\label{srb}
f_{+,3/2} f_{-,1/2} \frac{\eta^b_0}{\bar{\Lambda}} e^{-(\bar\Lambda+\bar{\Lambda}')/T}
  &=&\frac{1}{8\pi^2}  \frac{1+4y}{(1+y)^4} \int^{\omega^{*(1)}_c}_0 d\omega_{+} 
  e^{-\omega_{+}/T} \omega_{+}^4 \nonumber\\
  &-&\frac{\alpha_s  \langle FF \rangle }{96 \pi} \frac{7-y}{(1+y)^3} T-\frac{2\alpha_s  \langle \bar{q}q \rangle  }{3\pi} 
  \frac{T^2}{(1+y)^2}={\cal SR}_{b} \nonumber \\
 \label{src}
f_{+,3/2} f_{-,1/2} \frac{\eta^c_0}{\bar{\Lambda}'} e^{-(\bar\Lambda+\bar{\Lambda}')/T}
  &=&\frac{3}{8\pi^2}  \frac{2+3y}{(1+y)^4} \int^{\omega^{*(1)}_c}_0 d\omega_{+} 
  e^{-\omega_{+}/T} \omega_{+}^4 \nonumber\\
  &+&\frac{\alpha_s  \langle FF \rangle }{96 \pi} \frac{9y+1}{(1+y)^3} T-\frac{2\alpha_s  \langle \bar{q}q \rangle }{3\pi} 
  \frac{(3+2y) T^2}{(1+y)^2}={\cal SR}_{c} 
\end{eqnarray}
In the numerical calculations, we take the following typical values ($\alpha_s=g^2_s/4\pi$) for 
the condensates
 \begin{eqnarray}
   && \langle \bar{q} q \rangle \approx -(0.23 \; \mbox{GeV})^3; \nonumber\\
   &&i \langle \bar{q} \sigma_{\alpha \beta} F^{\alpha \beta} q \rangle \approx -m^2_0  \langle \bar{q} q \rangle  \;
     \mbox{with} \; m^2_0=0.8 \mbox{GeV}^2; \nonumber\\
   &&\alpha_s  \langle FF \rangle \equiv \alpha_s  \langle F^a_{\alpha\beta} F^{\alpha \beta}_a \rangle \approx 0.04 \;\mbox{GeV}^4 .
 \end{eqnarray}
From Eqs.(\ref{sr1})-(\ref{srtau}), (\ref{srb}), one easily gets
\begin{eqnarray}
\label{srratio}
\tau&=&\frac{ {\cal SR}_{\tau} }{\sqrt{ {\cal SR}_{-,1/2} \times {\cal SR}_{+,3/2} } }, \nonumber \\
\frac{\eta^b_0}{\bar{\Lambda}}&=&\frac{ {\cal SR}_b }{\sqrt{ {\cal SR}_{-,1/2} \times {\cal SR}_{+,3/2}} },
 \nonumber \\
\frac{\eta^c_0}{\bar{\Lambda}'}&=&\frac{{\cal SR}_c }{\sqrt{ {\cal SR}_{-,1/2} \times {\cal SR}_{+,3/2}} }.
\end{eqnarray}

The QCD higher order radiative corrections have not been included in 
eqs.(\ref{sr1})-(\ref{srtau}) and(\ref{srb}). But what we will use in our numerical analysis is 
eqs.(\ref{srratio}), which are ratios of the three-point correlators to the two-point 
correlators. Though the QCD radiative corrections may be large, one may expect that they 
may not influence the ratios in eqs.(\ref{srratio}) significantly due to the 
cancelation of numerators and denomenators. 

\section{Numerical analysis of wave functions and decay rates}\label{analysis}

Now we turn to the numerical analysis of the sum rules obtained in the previous section. 
Imposing usual criterium that both higher order power corrections and the contributions 
of the contimuun should not be very large, we find the proper sum rule "windows": 
$0.7\mbox{GeV} \langle T \langle 1.2\mbox{GeV}$. $f_{-,1/2}$ and $f_{+,3/2}$ have been studied in \cite{ww,neu1076,dh} 
and the corresponding thresholds were found to be in the ranges 
$1.6\mbox{GeV} \langle \omega_c \langle 2.2\mbox{GeV}$ and 
$2.7\mbox{GeV} \langle \omega^{**}_c \langle 3.2\mbox{GeV}$. In Fig.2 we present $\tau(1)$, 
$\eta^b_0(1)$ and $\eta^c_0(1)$ as functions of $T$ at fixed values 
of $\omega^*_c$ and $\omega^{*(1)}_c$, where $\omega_c=1.9$GeV and $\omega^{**}_c=2.95$GeV 
are used. We find that $\tau(1)$ are stable around the threshold $\omega^*_c \approx 2.35$GeV, 
whereas $\eta^b_0(1)$ and $\eta^c_0(1)$ are stable around a smaller threshold 
value $\omega^{*(1)}_c \approx 1.85$GeV. With this analysis, we obtain the following values for the wave 
functions 
\begin{eqnarray}
\tau(1)&=&0.8\pm 0.1 \mbox{GeV} \nonumber\\
-\frac{\eta^b_0(1)}{\bar{\Lambda}}&=&0.35\pm 0.04 \mbox{GeV} \nonumber\\
-\frac{\eta^c_0(1)}{\bar{\Lambda}'}&=&1.15\pm 0.15 \mbox{GeV} 
\end{eqnarray}
Where the errors mainly arise from the threshold. 
In Fig.3, the variations of $\tau$, $\eta^b_0$ and $\eta^c_0$ with respect to $y$ 
are shown, where we have used $T=1$GeV. 

We now come to consider the $B\to (D_1,D^*_2)$ decay rates. The differential 
decay rates are given by 
\begin{eqnarray}
\frac{d\Gamma(B\to D_1 l \bar\nu)}{dy}&=&\frac{G^2_F |V_{cb}|^2 m^5_B }{48 \pi^3} r^3_1 
   \sqrt{y^2-1} \{ 2(1-2y r_1+r^2_1) [f^2_{V_1}+(y^2-1)f^2_A ]\nonumber\\
   &+&[(y-r_1)f_{V_1}+(y^2-1)(f_{V_3}+r_1 f_{V_2}) ]^2 \}, \nonumber \\
\frac{d\Gamma(B\to D^*_2 l \bar\nu)}{dy}&=&\frac{G^2_F |V_{cb}|^2 m^5_B }{144 \pi^3} r^3_2 
   (y^2-1)^{3/2} \{ 3(1-2y r_2+r^2_2) [k^2_{A_1}+(y^2-1)k^2_V ]\nonumber\\
   &+&2 [(y-r_2)k_{A_1}+(y^2-1)(k_{A_3}+r_2 k_{A_2}) ]^2 \}
\end{eqnarray}
with the kinematically allowed ranges $1 \langle y \langle 1.32$ for $B\to D_1 l \bar\nu$ and 
$1 \langle y \langle 1.31$ for $B\to D^*_2 l \bar\nu$. 
The form factors $f_i$, $k_i$ are related to the wave 
functions as shown in (\ref{factorfunction}). The zero recoil values of $\kappa_1$ and 
$\kappa_2$ in (\ref{factorfunction}) have been evaluated 
by fitting from the ground state meson masses \cite{wwy1} and also by the QCD sum rule 
method \cite{ww}. Similarly, $\kappa'_i(1)$ can also be extracted from fitting the meson 
masses given in (\ref{lamb1rel}), i.e., 
\begin{equation}
\label{fitkappa21}
\kappa'_2(1)=\frac{m_c}{8} (m_{D^*_2}-m_{D_1}). 
\end{equation}
To extrat $\kappa'_1(1)$, we consider the spin average mass of each doublet:
\begin{equation}
\label{averagemass}
 \bar{m}_H=\frac{n_- m_{H-}+n_+ m_{H+} }{n_-+n_+}   
 \end{equation}
with 
\[ n_{\pm}=2 j_{\pm}+1, \;\;\; j_{\pm}=j_l \pm \frac{1}{2}  .   \]
(\ref{lamb1rel}), (\ref{lambrel}) and (\ref{averagemass}) yield
\begin{eqnarray}
\label{fitkappa11}
\kappa'_1(1)-\kappa_1(1)&=&\frac{[(\bar{m}'_B-\bar{m}_B)-(\bar{m}'_D-\bar{m}_D) ] m_b m_c}
    {m_b-m_c} , \nonumber \\
\bar{\Lambda}'-\bar{\Lambda}&=&\frac{ m_b (\bar{m}'_B-\bar{m}_B)-m_c (\bar{m}'_D-\bar{m}_D)}
    {m_b-m_c} .
\end{eqnarray}
Here $\bar{m}_{B(D)}$ are the average masses of the $j^P_l=\frac{1}{2}^-$ ground state 
mesons, whereas $\bar{m}'_{B(D)}$ are the ones of the $j^P_l=\frac{3}{2}^+$ doublet mesons. 

When taking the average meson masses to be  
$\bar{m}_D=1.971$GeV, $\bar{m}'_D=2.445$GeV, $\bar{m}_B=5.314$GeV and 
$\bar{m}'_B=5.73$GeV  \cite{akl} and the quark masses 
$m_b=4.8$GeV, $m_c=1.35$GeV, we arrive at the following values 
\begin{eqnarray}
\kappa_1(1)&=&-0.56 \mbox{GeV}^2 \nonumber \\
\kappa_2(1)&=&0.05 \mbox{GeV}^2 \nonumber \\
\kappa'_1(1)&=&-0.83 \mbox{GeV}^2 \nonumber \\
\kappa'_2(1)&=&0.00675 \mbox{GeV}^2 
\end{eqnarray}

$\eta^{Q}_i$ characterize the matrix elements of the chromomagnetic operator. They 
are often neglected from the argument that the chromomagnetic operator must have 
small effects due to the small $D^*_2-D_1$ mass splitting\cite{akl}.
In this section, we first neglect them but discuss their possible sizable effects late on. 
When $\eta^Q_i$ are neglected, the formulae of decay rates become very simple because each 
differential decay rate depends only on a composite wave function.
\begin{eqnarray}
\frac{d\Gamma(B\to D_1 l \bar\nu)}{dy}&=&\frac{G^2_F |V_{cb}|^2 m^5_B }{72 \pi^3} r^3_1 
   (y+1)(y^2-1)^{3/2} [ (y-1)(1+r_1)^2+y(1-2yr_1+r^2_1)] \tilde{\tau}^2_1(y), \nonumber\\
\frac{d\Gamma(B\to D^*_2 l \bar\nu)}{dy}&=&\frac{G^2_F |V_{cb}|^2 m^5_B }{72 \pi^3} r^3_2 
   (y+1)(y^2-1)^{3/2} [ (y+1)(1-r_2)^2+3y(1-2yr_2+r^2_2)] \tilde{\tau}^2_2(y)
\end{eqnarray}
with 
\begin{eqnarray}
\tilde{\tau}_1(y)&=&\tau-\frac{\eta^b_0}{2m_b\bar{\Lambda}}-\frac{\eta^c_0}{2m_c\bar{\Lambda}'}
  +\frac{\tau}{2m_b\bar{\Lambda}}(\kappa_1(1)+3\kappa_2(1))+\frac{\tau}{2m_c\bar{\Lambda}'}
  (\kappa'_1(1)+5\kappa'_2(1)), \nonumber \\
\tilde{\tau}_2(y)&=&\tau-\frac{\eta^b_0}{2m_b\bar{\Lambda}}-\frac{\eta^c_0}{2m_c\bar{\Lambda}'}
  +\frac{\tau}{2m_b\bar{\Lambda}}(\kappa_1(1)+3\kappa_2(1))+\frac{\tau}{2m_c\bar{\Lambda}'}
  (\kappa'_1(1)-3\kappa'_2(1)).
\end{eqnarray}
It is seen that the $1/m_Q$ corrections from the kinematic term do not change the relative 
values for the two differential decay rates. As the spin-symmetry breaking term $\kappa'_2$ arising 
from the mass splitting is small, without including the $1/m_Q$ corrections from the chromomagnetic
terms, the relative value of their total decay rates should have the similar behavior as the one 
at the leading order, i.e., 
\begin{equation}
R = \frac{\Gamma^0(B\to D_1 l \bar\nu)}{\Gamma^1(B\to D_1 l \bar\nu)} \sim 
\frac{\Gamma^0(B\to D_2^* l \bar\nu)}{\Gamma^1(B\to D_2^* l \bar\nu)}
\end{equation}
which can be explicitly seen from Table 1, where we have used the input values   
$m_b=4.8$GeV, $m_c=1.35$GeV, $|V_{cb}|=0.038$, the life time $\tau_B=1.6$ps 
and the thresholds $\omega_c=1.9$GeV, $\omega^{**}_c=2.95$GeV, 
$\omega^*_c=2.35$GeV, $\omega^{*(1)}_c=1.85$GeV.
Note that both decay rates of $B\to D_1 l \bar\nu$ and $B\to D^*_2 l \bar\nu$ receive large 
$1/m_Q$ contributions (but not as large as the $1/m_Q$ contributions received by 
the $B\to D_1 l \bar\nu$ decay rate given in \cite{ebert} within the framework of the 
relativistic quark model based on the quasipotential approach. 
This is different from the discussions based on the quasipotential approach with the same structure 
of heavy quark mass corrections predicted in the usual HQET \cite{ebert}, where the decay rate of 
$B\to D^*_2 l \bar\nu$ is only slightly increased by subleading $1/m_Q$ corrections. 

In comparison with the experimental data reported by CLEO \cite{cleo} and ALEPH \cite{aleph} groups, 
our result for the branching ratio of the $B\to D_1 l \bar\nu$ decay with the inclusion of 
$1/m_Q$ corrections is in agreement with both measurements. While for the $B\to D^*_2 l \bar\nu$ 
decay, the result at $m_Q\to \infty$ limit is within the CLEO upper limit but not 
within the ALEPH one. When including $1/m_Q$ contributions, we find that the resulting decay rate for
$Br(B\to D^*_2 l \bar\nu)$ seems to go over the CLEO upper limit though it may still be consistent within
the large erros due to the big uncertainties of the choices of the thresholds. On the other hand, 
in deriving the results in Table 1, we have made the assumption that $\eta^Q_i=0$.
In general, the contributions of $\eta^Q_i$ may not be neglected. Their effects have shown in Fig.4, 
where we have used the above sum rule results for $\tau$, $\eta^b_0$ and $\eta^c_0$ and 
the thresholds $\omega_c=1.9$GeV, $\omega^{**}_c=2.95$GeV, $\omega^*_c=2.35$GeV and 
$\omega^{*(1)}_c=1.85$GeV. We have also made the assumption that $\eta^Q_i$ and $\tilde{\tau}$ have 
the similar dependence on $y$, $\tilde\tau \approx \tilde\tau(1)/(1+\frac{y-1}{a^2})$ with 
$a^2 \approx 0.7 $. It is seen from Fig.4 
that $\eta^c_1$, $\eta^c_2$, $\eta^c_3$ and $\eta^b$ influence the 
branching ratios in different ways. When $\eta^c_1$  becomes larger, 
 $Br(B\to D_1 l \bar\nu)$ decreases while $Br(B\to D^*_2 l \bar\nu)$ 
increases, and $\eta^c_2$ influences the two branching ratios in the same manner 
as $\eta^c_1$ does, but both branching ratios are not sensitive to $\eta^c_2$. 
$\eta^c_3$ has opposite effects on the two decay modes, i.e., its increment enlarges 
$Br(B\to D_1 l \bar\nu)$ but suppresses $Br(B\to D^*_2 l \bar\nu)$. For $\eta^b$, 
it always appears in the form $\tilde{\tau}+\frac{\eta^b}{m_b \bar{\Lambda}}$, so it affects 
$Br(B\to D_1 l \bar\nu)$ and $Br(B\to D^*_2 l \bar\nu)$ in the same way, i.e., a negative 
value of $\eta^b$ may suppress both branching ratios. 
Some reasonable results for the branching ratios at certain values of $\eta^Q_i$ are listed in Table 2 
and Table 3.
From Table 2, Table 3 and Fig.4, we see that both 
branching ratios of $B\to D_1 (D^*_2) l \bar\nu$ may be suppressed after considering the possible 
effects due to the contributions of chromomagnetic terms at $1/m_Q$ order. The resulting two branching ratios 
can easily be made to be consistent with the experimental measurements when $\eta^b$, $\eta^c_1$, $\eta^c_2$ 
and $\eta^c_3$ are in the proper ranges.  

 \section{Summary}\label{sum}

The HQEFT has been reviewed and applied to study the semileptonic $B$ decays 
into excited charmed mesons $(D_1, D^*_2)$ with $j_l^P=\frac{3}{2}^+$. The form factors 
of the matrix elements relevant to these two decays have been expressed in terms of wave functions 
within the framework of HQEFT. It has been shown that with the inclusion of quark-antiquark coupled sectors, 
the relevant matrix elements can be parameterized by the leading order Isgur-Wise function and 
additional twelve wave functions $\eta^Q_i\;(Q=b,c;\;i=0,1,2,3)$ and $\kappa^{(')}_j\;(j=1,2)$ 
appearing at $1/m_Q$ order. Two wave functions $\tau^b_1$ and $\tau^c_1$ characterizing
 the $1/m_Q$ corrections of effective current in the usual HQET \cite{aklz,akl}
are absent in HQEFT. 

 By adopting proper interpolating currents for the 
heavy-light mesons, the leading order wave function $\tau$ and two important wave functions 
$\eta^b_0$ and $\eta^c_0$ at $1/m_Q$ order have been calculated by the QCD sum rule 
approach. Zero recoil values of $\kappa'_1$ and $\kappa'_2$, which are $1/m_Q$ order 
wave functions for the matrix elements between $\frac{3}{2}^+$ excited states, 
have also been extracted from fitting the excited meson masses. 
By using these calculated results and also considering possible effects 
from wave functions $\eta^Q_i \;(i=1,2,3)$ arising from the chromomagnetic operators at $1/m_Q$ order, 
$B\to D_1(D^*_2) l \bar\nu$ decay rates and branching ratios 
have been evaluated. When neglecting $\eta^Q_i \;(i=1,2,3)$, we have shown 
that both of these decays receive large $1/m_Q$ corrections from $\eta^Q_0$ and 
$\kappa^{(')}_i(1)\;(i=1,2)$. For $B\to D_1l\bar\nu$ decay this is similar to the results 
obtained from quasiquark potential approach based on the same structure as in the usual HQET, 
but for $B\to D^*_2l\nu$ decay the situation is quite different from the case described by that 
approach, where $Br(B\to D^*_2l\bar\nu)$ only increases slightly when the $1/m_Q$ order contributions 
from $\tau^{b(c)}_1$ are included. It has been shown that when considering all possible contributions 
from wave functions at the $1/m_Q$ order in the HQEFT, but without breaking the $1/m_Q$ expansion, 
the resulting branching ratios of the semileptonic $B$ decays into $\frac{3}{2}^+$ excited 
charmed mesons  can agree well with the experimental 
measurements for the proper ranges of $\eta^Q_i$. 

\acknowledgments

We would like to thank Y. B. Dai and M. Q. Huang for 
useful discussions. This work was supported in part by the NSF of China under the grant No. 19625514.

\appendix

 \section{Locality of HQEFT}\label{HQEFT}

For completeness and clarification, we briefly review the formulation of heavy quark effective 
field Lagrangian that keeps both effective quark and antiquark fields\cite{ylw}. In particular, 
we would like to emphasize, by adding this appendix, that the HQEFT is a local effective theory and 
contains no non-local operators. 

 Firstly,  denote the heavy quark (antiquark) field as
     \begin{equation}
     \label{eq:qzf}
       Q=Q^{+}+Q^{-} ,
     \end{equation}
  with $Q^{+}$ and $Q^{-}$ corresponding to the two solutions of the Dirac equation.
  
  Defining
    \begin{equation}
    \label{eq:quarkfield}
      \hat{Q}^{\pm}_v\equiv \frac{1{\pm}v\hspace{-0.2cm}\slash}{2}Q^{\pm},\hspace{1.5cm}
      R^{\pm}_v\equiv \frac{1{\mp}v\hspace{-0.2cm}\slash}{2}Q^{\pm}
    \end{equation} 
  with $v^{\mu}$ an arbitrary four-vector satisfying $v^2=1$. Furthermore $\hat{Q}^{\pm}_v$ 
  are related to the desired effective fields defined as 
\begin{eqnarray}
    \label{eq:qv}
      Q_v=e^{i v\hspace{-0.15cm}\slash m_Qv\cdot x}\hat{Q}_v, \;\;\;\; 
      \bar{Q}_v=\bar{\hat{Q}}_v e^{-iv\hspace{-0.15cm}\slash m_Qv\cdot x}.
\end{eqnarray}     
 From the equation of motion of quark field and antiquark field, the original fields 
 $Q$ and $\bar{Q}$ can be expressed by the new fields as follows,
  \begin{eqnarray}
   \label{eq:Qexpre}
       \left\{
        \begin{array}{l}
  Q^{\pm}=[1+(1-\frac{i\DSPX+\MQ}{2\MQ})^{-1}
           \frac{i\DSCX}{2\MQ}] \hat{Q}^{\pm}_{v}\equiv\hat{\omega} \hat{Q}^{\pm}_v \\
         \hspace{0.7cm} =e^{\mp im_Q v\cdot x} [1+(1-\frac{i\DSPX}{2m_Q})^{-1} \frac{i\DSC}{2m_Q}]Q^{\pm}_v
          \equiv e^{\mp im_Q v\cdot x} \omega Q^{\pm}_v ,\\
  \bar{Q}^{\pm} = \bar{\hat{Q}}^{\pm}_{v}[1+\frac{-i\stackrel{\hspace{-0.1cm}\leftarrow}
         {\DSCX}}{2\MQ}(1-\frac{-i\stackrel{\hspace{-0.1cm}\leftarrow}
         \DSPX+\MQ}{2\MQ})^{-1}]\equiv \bar{\hat{Q}}^{\pm}_v\stackrel{\leftarrow}
           {\hat{\omega}} \\
       \hspace{0.7cm} =\bar{Q}^{\pm}_{v}[1+\frac{-i\stackrel{\hspace{-0.1cm}\leftarrow}
         {\DSCX}}{2\MQ}(1-\frac{-i\stackrel{\hspace{-0.1cm}\leftarrow}
         \DSPX}{2\MQ})^{-1}] e^{\pm im_Q v\cdot x}
        \equiv \bar{Q}^{\pm}_v\stackrel{\leftarrow}
           {\omega} e^{\pm im_Q v\cdot x}     .
         \end{array}
        \right.  
    \end{eqnarray}
  $\stackrel{\hspace{-0.1cm}\leftarrow}{D^\mu}$, $\DSP$ and $\DSC$ are 
  defined as
      \begin{equation}
      \label{eq:12}
       \left\{
        \begin{array}{l}
           \int\kappa\stackrel{\hspace{-0.1cm}\leftarrow}{D^\mu}\varphi 
   \equiv -\int\kappa D^{\mu}\varphi, \\
           \DSP \equiv v\hspace{-0.2cm}\slash(v\cdot D), \\
           \DSC \equiv \DS-\VS (v\cdot D).
       \end{array}
         \right.
      \end{equation}
  The QCD Lagrangian becomes
    \begin{equation}
    \label{eq:LQCD}
       {\cal L}_{QCD} = {\cal L}_{light}+{\cal L}_{eff} ,
    \end{equation}
  where ${\cal L}_{light}$ represents the part of Lagrangian containing no heavy 
  quarks, and 
     \begin{equation}
     \label{eq:LQV}
        {\cal L}_{eff} = {\cal L}^{++}_{v}+{\cal L}^{+-}_{v}+{\cal L}^{-+}_{v}+{\cal L}^{--}_{v}
     \end{equation}
  with 
    \begin{eqnarray}
    \label{eq:LZF}
   {\cal L}^{\pm \pm}_{v} &=& \QVHPMB [i\DSP-\MQ 
           + \frac{1}{2\MQ}i\DSC (1-\frac{i\DSP+\MQ}{2\MQ})^{-1}
           i\DSC ] \QVHPM \equiv  \QVHPMB \hat{A} \QVHPM \nonumber \\ 
      &=& \bar{Q}^{\pm}_v [i\DSP+\frac{1}{2\MQ}i\DSC (1-\frac{i\DSP}{2\MQ})^{-1}
           i\DSC ] Q^{\pm}_v \equiv \bar{Q}^{\pm}_v A Q^{\pm}_v \nonumber \\ 
   {\cal L}^{\pm \mp}_{v} &=& \QVHPMB [ -i\DSC+\frac{1}{4\MQ^2}
         (-i\stackrel{\hspace{-0.2cm}\leftarrow}{\DSC}) 
       (1-\frac{-i\stackrel{\hspace{-0.1cm}\leftarrow}{\DSP}+\MQ}{2\MQ})^{-1}i\DSC \nonumber \\
    &\times&     (1-\frac{i\DSP+\MQ}{2\MQ})^{-1} i\DSC ] \QVHMP 
        \equiv \QVHPMB \hat{B}  \QVHMP \nonumber \\
      &=& e^{\pm 2im_Q v\cdot x} \bar{Q}^{\pm}_v [ -i\DSC+\frac{1}{4\MQ^2}
         (-i\stackrel{\hspace{-0.2cm}\leftarrow}{\DSC}) 
          (1-\frac{-i\stackrel{\hspace{-0.1cm}\leftarrow}{\DSP}}{2\MQ})
         ^{-1}i\DSC \nonumber \\
      &\times& (1-\frac{i\DSP}{2\MQ})^{-1} i\DSC ]  Q^{\mp}_v
      \equiv e^{\pm 2im_Q v\cdot x} \bar{Q}^{\pm}_v B Q^{\mp}_v 
    \end{eqnarray}
It is seen that all parts of the effective Lagrangian are local.   

  When quark fields and antiquark fields decouple completely, it is reasonable 
  to deal with only section ${\cal L}^{++}_{v}$ or ${\cal L}^{--}_{v}$ independently. 
  This is just the case considered in the framework of the usual HQET. 
  To consider the finite quark mass corrections one should also include the 
  contributions from ${\cal L}^{\pm\mp}_{v}$ and ${\cal L}^{--}_{v}$ as well.

Similarly, the heavy quark currents can also be decomposed into four parts in a similar way, 
\begin{eqnarray}
\label{eq:Jexpand}
 J(x)&=&\bar{Q}^{\prime}(x)\Gamma Q(x) 
  =\bar{Q}^{\prime +}\Gamma Q^{+}+\bar{Q}^{\prime +}\Gamma Q^{-}
     +\bar{Q}^{\prime -}\Gamma Q^{+}+\bar{Q}^{\prime -}\Gamma Q^{-} \nonumber \\
 \to J_{eff}(x) &=& \QVPHZB \stackrel{\leftarrow}{\hat\omega} \Gamma \hat\omega \QVHZ 
    + \QVPHZB \stackrel{\leftarrow}{\hat\omega} \Gamma \hat\omega \QVHF 
    + \QVPHFB \stackrel{\leftarrow}{\hat\omega} \Gamma \hat\omega \QVHZ 
    + \QVPHFB \stackrel{\leftarrow}{\hat\omega} \Gamma \hat\omega \QVHF \nonumber \\
  &=& J^{++}_v+J^{+-}_v+J^{-+}_v+J^{--}_v.
\end{eqnarray}
 
The matrix element on the r.h.s. of (\ref{eq:statedef}) is actually evaluated via 
\begin{eqnarray}
 \langle H'_{v'}| J_{eff} e^{i\int {\cal L}_{eff} } |H_v \rangle = \langle H'_{v'}| (J^{++}_v+J^{+-}_v+J^{-+}_v+J^{--}_v) 
 e^{i\int ({\cal L}^{++}_v+{\cal L}^{+-}_v+{\cal L}^{-+}_v+{\cal L}^{--}_v)} |H_v \rangle .
\end{eqnarray}
Here $J_{eff}$ and ${\cal{L}}_{eff}$ shall include all 4 parts instead of only the $`++'$ parts 
in (\ref{eq:LQV}) and (\ref{eq:Jexpand}) as has been done in the usual HQET. 
The first sector $`++'$ of $J_{eff}$ contributes
\begin{eqnarray}
 \langle H'_{v'}| \QVPHZB \stackrel{\leftarrow}{\hat{\omega}} \Gamma \hat\omega \QVHZ 
  e^{i\int {\cal{L}}_{eff} } | H_v \rangle .
\end{eqnarray}
If neglecting all except $`++'$ sectors in $J_{eff}$ and ${\cal L}_{eff}$, we get 
\[  \langle H'_{v'}|\QVPHZB \stackrel{\leftarrow}{\hat{\omega}} \Gamma \hat\omega \QVHZ 
  e^{i\int \QVHZB \hat{A} \QVHZ } | H_v \rangle ,\]
which is just what the usual HQET treated. 

The contributions from the $`+-'$ sector of $J_{eff}$ is 
\begin{equation}
\label{matzf}
 \langle H'_{v'}|\QVPHZB \stackrel{\leftarrow}{\hat{\omega}} \Gamma \hat\omega \QVHF 
  e^{i\int {\cal L}_{eff} } | H_v \rangle .
\end{equation} 
In the case that both the initial and final states contain only heavy quarks (no 
heavy antiquarks), matrix elements such as 
\[ \langle H'_{v'}|\QVPHZB \stackrel{\leftarrow}{\hat{\omega}} \Gamma \hat\omega \QVHF| H_v \rangle \] 
do not contribute, and (\ref{matzf}) contributes only at higher perturbation order, i.e., 
only when ${\cal L}_{eff}$ is inserted into the matrix elements. Therefore the leading 
order contribution from (\ref{matzf}) should be 
\begin{equation}
\label{QVZF}
 \langle H'_{v'}| i\int dx dy T\{ (\QVPHZB \stackrel{\leftarrow}{\hat{\omega}} \Gamma \hat\omega \QVHF)(x), 
 (\QVHFB \hat{B} \QVHZ)(y) \}| H_v \rangle 
\end{equation} 
and then the effective antiquark fields $\QVHF(x)$ and $\QVHFB(y)$ should be contracted. 
Due to (\ref{eq:LZF}), this will yield the propagator 
\[ \frac{-iP_- }{\VS v\cdot p+m_Q+O(1/m_Q) } \;\;\; \; (P_{\pm}\equiv \frac{1\pm \VS}{2}),\]
which is of $O(1/m_Q)$. 

Contributions from other sectors of the current $J_{eff}$ can be treated in the same 
way. It is easy to see that $ \langle H'_{v'}| J^{++}_v e^{i\int {\cal L}_{eff}} | H_v \rangle $ gives contributions 
of $O(1)$, and $ \langle H'_{v'}| J^{\pm \mp}_v e^{i\int {\cal L}_{eff}} | H_v \rangle $ is $O(1/m_Q)$, 
whereas $ \langle H'_{v'}| J^{--}_v e^{i\int {\cal L}_{eff}} | H_v \rangle $ is $O(1/m^2_Q)$ since 
${\cal L}_{1/m_Q}$ should be inserted twice, and each contraction of $\hat{Q}^{-}_v$ and 
$\QVHFB$ gives a $1/m_Q$ suppression. 

To be more clear, contracting $\QVHFB$ and $\QVHF$ in (\ref{QVZF}) yields 
\begin{eqnarray} 
 \langle H'_{v'}| i\int d^4x d^4y \QVPHZB(x) \stackrel{\leftarrow}{\hat\omega} \Gamma \hat\omega 
  \int \frac{d^4p}{(2\pi)^4} \frac{-iP_-\ e^{-ip\cdot (x-y)} }{\VS v\cdot p+m_Q+O(1/m_Q)} 
  \hat{B} \QVHZ (y) e^{i\int {\cal L}_{eff} } |H_v \rangle .
\end{eqnarray}
which can be written in the following form with replacing the momentum $p$ by the derivative $\partial$ 
and performing the integral of the momentum   
\begin{eqnarray} 
 \langle H'_{v'}| i\int d^4x d^4y \QVPHZB(x) \stackrel{\leftarrow}{\hat\omega} \Gamma \delta(x-y) 
  \hat\omega \frac{P_-}{i\VS v\cdot \partial  +m_Q+O(1/m_Q) } \hat{B} \QVHZ(y) e^{i\int {\cal L}_{eff} } 
  |H_v \rangle ,
\end{eqnarray} 
Using the same trick, one can treat all contributions from $ \langle H'_{v'}|J^{+-}_v e^{i\int {\cal L}_{eff}}|H_v \rangle $ 
and obtains
\begin{eqnarray}
& &  \langle H'_{v'}|J^{+-}_v e^{i\int {\cal L}_{eff}}|H_v \rangle  \nonumber \\
& & =  \langle H'_{v'}| \QVBP \stackrel{\leftarrow}{\omega} 
 \Gamma \omega [-i\DSP-\frac{1}{2m_Q} i\DSC (1-\frac{i\DSP}{2m_Q})^{-1} i\DSC ]^{-1} 
 B \QV e^{i\int {\cal L}_{eff} } |H_v \rangle  \nonumber \\
& & =  \langle H'_{v'}| e^{i(m_{Q'}v'-m_Q v)\cdot x} \QVBP \stackrel{\leftarrow}{\omega} 
  \Gamma \omega (-A^{-1})B \QV  e^{i\int {\cal L}_{eff}} |H_v \rangle 
\end{eqnarray} 
which means that, effectively, one can reexpress  $J^{+-}$ to be the effective current in which only the 
effective quark fields $\QVBP$ and $\QV$ are used. Namely, 
\begin{eqnarray}
J^{+-}_v&=& \langle Q^{'+} \Gamma Q^- \rangle  =\QVPHZB \stackrel{\leftarrow}{\hat{\omega}}
  \Gamma \hat\omega \QVHF \nonumber \\
&\to& \QVPHZB \stackrel{\leftarrow}{\hat{\omega}}\Gamma \hat\omega (-\hat{A}^{-1}) \hat{B} \QVHZ 
  =e^{i(m'_{Q'}v'-m_Q v)\cdot x} 
  \QVBP \stackrel{\leftarrow}{\omega}\Gamma \omega (-A^{-1}) B \QV 
\end{eqnarray}

In an analogous way, one can reexpress $J^{-+}_v$, $J^{--}_v$, ${\cal L}^{+-}_v$, ${\cal L}^{-+}_v$ and 
${\cal L}^{--}_v$ into the corresponding effective currents and Lagrangians by only using  
the effective quark fields $\QVBP$ and $\QV$. Consequently, we have 
 \begin{eqnarray}
    \label{eq:oth}
{\cal L}&\to& {\cal L}_{eff}\to {\cal L}_{eff}^{++} \equiv {\cal L}^{++}_v+\tilde{\cal L}^{++}_v, \\
\tilde{\cal L}^{++}_{v} & \equiv&  \langle {\cal L}^{+-}_v+{\cal L}^{-+}_v+{\cal L}^{--}_v \rangle  
 = \frac{1}{2m_Q}\QVB  [i\DSP+\frac{1}{2m_Q} i\DSC (1-\frac{i\DSP}{2m_Q})^{-1}i\DSC 
     ]\frac{1}{i\DSP} i\DSC \nonumber \\
& \times &  [1-\frac{i\DSP}{2m_Q}+\frac{1}{2m_Q}i\DSC \frac{1}{i\DSP}i\DSC ]^{-1}
  i\DSC \frac{1}{i\DSP}  [i\DSP+\frac{1}{2m_Q} i\DSC (1-\frac{i\DSP}{2m_Q})^{-1} 
  i\DSC ]  \QV  \nonumber\\ & =& \frac{1}{2m_Q}\QVB A \frac{1}{i\DSP} i\DSC  
[1-\frac{i\DSP}{2m_Q}+\frac{1}{2m_Q}i\DSC \frac{1}{i\DSP}i\DSC ]^{-1}
  i\DSC \frac{1}{i\DSP} A \QV ,
 \end{eqnarray}  
which represends the additional contributions to the effective Lagrangian ${\cal L}^{++}_v$ that
has been widely adopted in the usual HQET.  This additional part may be regarded as the effective
potential part of heavy quark due to the exchanges of virtual antiquarks.
It is seen that when one imposes the on-shell condition $AQ^+_v=0$, i.e.
 ${\cal L}^{++}_v=0$, the effective potential part $\tilde{\cal L}^{++}_v$ also vanishes,
i.e. $\tilde{\cal L}^{++}_v=0$. For off-shell case $ \langle i\DSP \rangle \sim  \langle i\DSC \rangle \sim \bar\Lambda$,or
 \[\frac{ \langle (i\DSCX)^2 \rangle }{2m_Q}\  \ll \   \langle i\DSP \rangle ,\] the leading contribution of the effective 
potential part is 
\[  \tilde{\cal L}^{++}_v|_{LO} =\QVB \frac{(i\DSC)^2}{2m_Q} \QV   .     \]
Correspondingly, the heavy quark current turns out to be \cite{wwy1,ww,wwy2,wy} 
 \begin{eqnarray}
   \label{eq:Jeff}
   J&=& \bar{Q}'\Gamma Q \to J_{eff}\to J_{eff}^{++} \equiv J^{++}_v+\tilde{J}^{++}_v  \nonumber \\  
      & = & e^{i(\hat{m}_{Q^{\prime}} v^{\prime}
        -\hat{m}_Q v)\cdot x}
       \{ \QVBP \Gamma \QV+\frac{1}{2\MQ}\QVBP\Gamma\frac{1}{i\DSP}(i\DSC)^2 \QV  \nonumber\\
        &&+\frac{1}{2\MQP}\QVBP (-i\stackrel{\hspace{-0.1cm}\leftarrow}{\DSC})^2\frac{1}
        {-i\stackrel{\hspace{-0.1cm}\leftarrow}{\DSP}}\Gamma\QV+\frac{1}{4\MQ^2}\QVBP 
        \Gamma\frac{1}{i\DSP}
        i\DSC (i\DSP)  \nonumber\\
        &&\times i\DSC \QV 
        +\frac{1}{4\MQP^2}\QVBP (-i\stackrel{\hspace{-0.1cm}\leftarrow}{\DSC})
        (-i\stackrel{\hspace{-0.1cm}\leftarrow}{\DSP})(-i\stackrel{\hspace{-0.1cm}\leftarrow}{\DSC})
        \frac{1}{-i\stackrel{\hspace{-0.1cm}\leftarrow}{\DSP}}\Gamma \QV \nonumber\\
        &&+\frac{1}{4\MQP \MQ}\QVBP (-i\stackrel{\hspace{-0.1cm}\leftarrow}{\DSC})^2 
        \times \frac{1}{-i\stackrel{\hspace{-0.1cm}\leftarrow}{\DSP}}\Gamma     
        \frac{1}{i\DSP}(i\DSC)^2 \QV
        +O(\frac{1}{m_{Q^{(\prime)}}^3}) \} \nonumber\\
        & \equiv & J^{(0)}_{eff}+J^{(1/m_Q)}_{eff}
    \end{eqnarray}
  with $J^{(0)}_{eff}$ the leading term $J^{(0)}_{eff}=e^{i(\hat{m}_{Q^{\prime}} v^{\prime}
  -\hat{m}_Q v)\cdot x}\QVBP \Gamma \QV $ and $J^{(1/m_Q)}_{eff}$ the remaining 
  terms in $J_{eff}$.

\section{Lorentz Structures}\label{Los}
Here we present the general Lorentz structures of $L^{\mu\nu(\mu\alpha\beta)}_{V,A} $ appeared in 
eqs.(\ref{correlator1})-(\ref{correlator6}).
\begin{eqnarray}
\label{lorentz}
 {\cal L}^{\mu\nu}_{V}&=&\frac{1}{\sqrt{6}}\left[(y^2-1)g^{\mu\nu}+3v^\mu v^\nu+
 (1-2y)v'^\mu v'^\nu-3yv^\mu v'^\nu-(y-2)v'^\mu v^\nu\right]\;,\\
 {\cal L}^{\mu\nu}_{A}&=&i\frac{1}{\sqrt{6}} (1+y)\epsilon ^{\mu\nu\alpha\beta}v_\alpha
 v'_\beta\;,\\
 {\cal L}^{\mu\alpha\beta}_{2V}&=&-\frac{i}{2}\left(\epsilon^{\mu\alpha\sigma\rho}
 (v_\beta-v'_\beta y)+\epsilon^{\mu\beta\sigma\rho}
 (v_\alpha-v'_\alpha y)\right)v_\sigma v'_\rho\;,\\
 {\cal L}^{\mu\alpha\beta}_{A}&=&\frac{1}{3}(1+y)g^{\alpha\beta}(v^\mu-v'^\mu)
 -\frac{1}{2}(1+y)g^{\alpha\mu}(v^\beta-v'^\beta y)
 -\frac{1}{2}(1+y)g^{\beta\mu}(v^\alpha-v'^\alpha y)\nonumber\\
    &&\quad\mbox{}
 +\frac{1}{2}(1-y)v'^\alpha v^\beta v'^\mu 
 +\frac{1}{2}(1-y)v^\alpha v'^\beta v'^\mu -\frac{1}{3}(1+y)
 v'^\alpha v'^\beta v^\mu\nonumber\\
    &&\quad\mbox{}
 -\frac{2}{3} (1+y)v'^\alpha v'^\beta v'^\mu
 +(v'^\alpha v'^\beta +v^\alpha v^\beta)v'^\mu.
\end{eqnarray}



\newpage
\centerline{\large{FIGURES}}

\small
\begin{center}
\begin{picture}(400,350)(0,0)
\SetWidth{2}
\Line(20,250)(50,300)
\Line(50,300)(80,250)
\Photon(50,320)(50,300){3}{2}
\SetWidth{0.5}
\Line(20,250)(80,250)
\BBoxc(50,300)(8,8)
\DashLine(0,250)(20,250){3}
\DashLine(80,250)(100,250){3}

\SetWidth{2}
\Line(170,250)(200,300)
\Line(200,300)(230,250)
\Photon(200,320)(200,300){3}{2}
\SetWidth{0.5}
\Line(170,250)(190,250) \Vertex(190,250){3}
\Line(210,250)(230,250) \Vertex(210,250){3}
\BBoxc(200,300)(8,8)
\DashLine(150,250)(170,250){3}
\DashLine(230,250)(250,250){3}

\SetWidth{2}
\Line(320,250)(350,300)
\Line(350,300)(380,250)
\Photon(350,320)(350,300){3}{2}
\SetWidth{0.5}
\Line(320,250)(340,250) \Vertex(340,250){3}
\Line(360,250)(380,250) \Vertex(360,250){3}
\GlueArc(343,250)(10,40,180){3}{1}
\Vertex(350,260){3}
\BBoxc(350,300)(8,8)
\DashLine(300,250)(320,250){3}
\DashLine(380,250)(400,250){3}

\SetWidth{2}
\Line(20,150)(50,200)
\Line(50,200)(80,150)
\Photon(50,220)(50,200){3}{2}
\SetWidth{0.5}
\Line(20,150)(40,150) \Vertex(40,150){3}
\Line(60,150)(80,150) \Vertex(60,150){3}
\GlueArc(57,150)(10,0,140){3}{1}
\Vertex(50,160){3}
\BBoxc(50,200)(8,8)
\DashLine(0,150)(20,150){3}
\DashLine(80,150)(100,150){3}

\SetWidth{2}
\Line(170,150)(200,200)
\Line(200,200)(230,150)
\Photon(200,220)(200,200){3}{2}
\SetWidth{0.5}
\Line(170,150)(190,150) \Vertex(190,150){3}
\Line(210,150)(230,150) \Vertex(210,150){3}
\GlueArc(200,145)(18,10,170){3}{3}
\BBoxc(200,200)(8,8)
\DashLine(150,150)(170,150){3}
\DashLine(230,150)(250,150){3}

\SetWidth{2}
\Line(320,150)(350,200)
\Line(350,200)(380,150)
\Photon(350,220)(350,200){3}{2}
\SetWidth{0.5}
\Line(320,150)(340,150) \Vertex(340,150){3}
\Line(360,150)(380,150) \Vertex(360,150){3}
 \GlueArc(347,138)(24,22,158){3}{3}
 
\BBoxc(350,200)(8,8)
\DashLine(300,150)(320,150){3}
\DashLine(380,150)(400,150){3}

\SetWidth{2}
\Line(20,50)(50,100)
\Line(50,100)(80,50)
\Photon(50,120)(50,100){3}{2}
\SetWidth{0.5}
\Line(20,50)(80,50)

 \Gluon(35,50)(40,65){3}{1}
\Vertex(40,65){3}
 \Gluon(60,65)(65,50){3}{1}
\Vertex(60,65){3}

\BBoxc(50,100)(8,8)
\DashLine(0,50)(20,50){3}
\DashLine(80,50)(100,50){3}

\SetWidth{2}
\Line(170,50)(200,100)
\Line(200,100)(230,50)
\Photon(200,120)(200,100){3}{2}
\SetWidth{0.5}
\Line(170,50)(230,50) 

 \Gluon(170,50)(190,65){3}{1}
\Vertex(190,65){3}
 \Gluon(210,65)(215,50){3}{1}
\Vertex(210,65){3}

\BBoxc(200,100)(8,8)
\DashLine(150,50)(170,50){3}
\DashLine(230,50)(250,50){3}

\SetWidth{2}
\Line(320,50)(350,100)
\Line(350,100)(380,50)
\Photon(350,120)(350,100){3}{2}
\SetWidth{0.5}
\Line(320,50)(340,50) \Vertex(340,50){3}
\Line(360,50)(380,50) \Vertex(360,50){3}
 \GlueArc(347,38)(24,90,158){3}{1}
\Vertex(350,60){3}
\BBoxc(350,100)(8,8)
\DashLine(300,50)(320,50){3}
\DashLine(380,50)(400,50){3}


\end{picture}
\end{center}

Fig.1. Feynman diagrams contributing to $\eta^b_0$ and $\eta^c_0$ at the order concerned. 
The box at the up of each diagram represents the $1/m_Q$ order heavy-heavy currents 
in eqs.(\ref{correlator1})-(\ref{correlator6}).

\newcommand{\PIC}[2]
{
\begin{center}
\begin{picture}(300,300)(0,0)
\put(40,25){
\epsfxsize=8cm
\epsfysize=8cm
\epsffile{#1} }
\put(150,40){\makebox(0,0){#2}}
\end{picture}
\end{center}
}

\newpage
\small
\mbox{}

\PIC{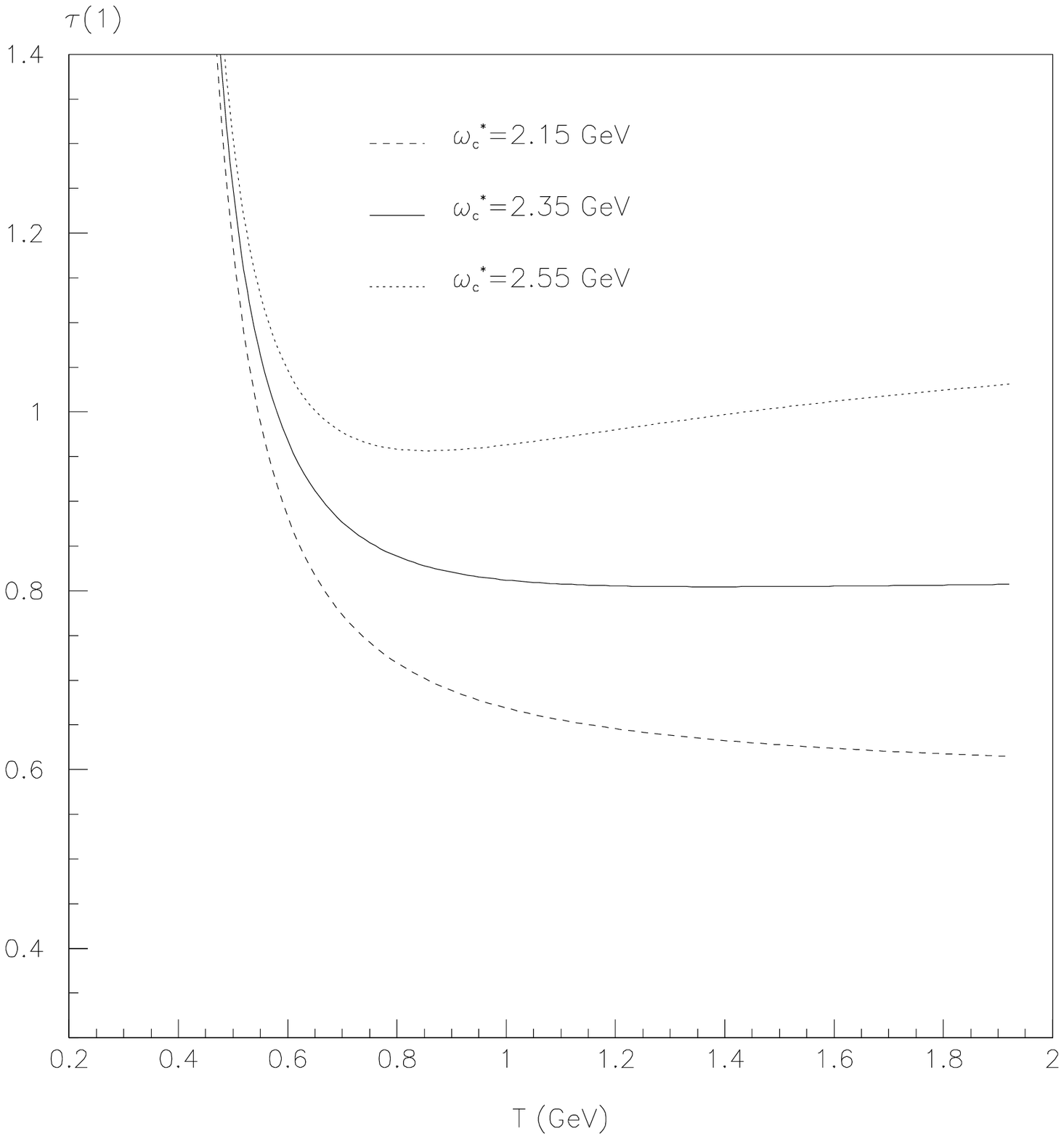}{Fig.2a}

\PIC{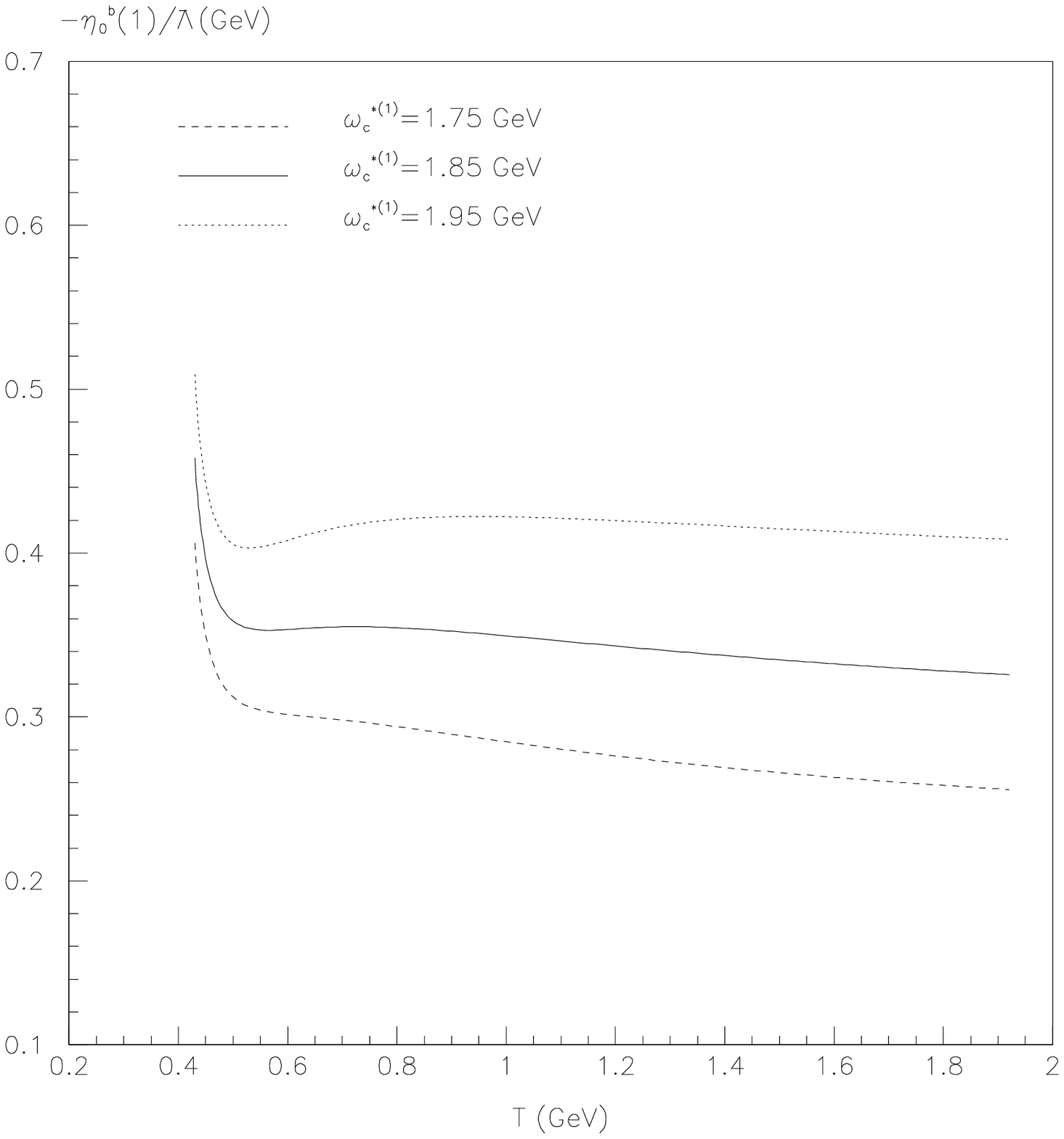}{Fig.2b}

\PIC{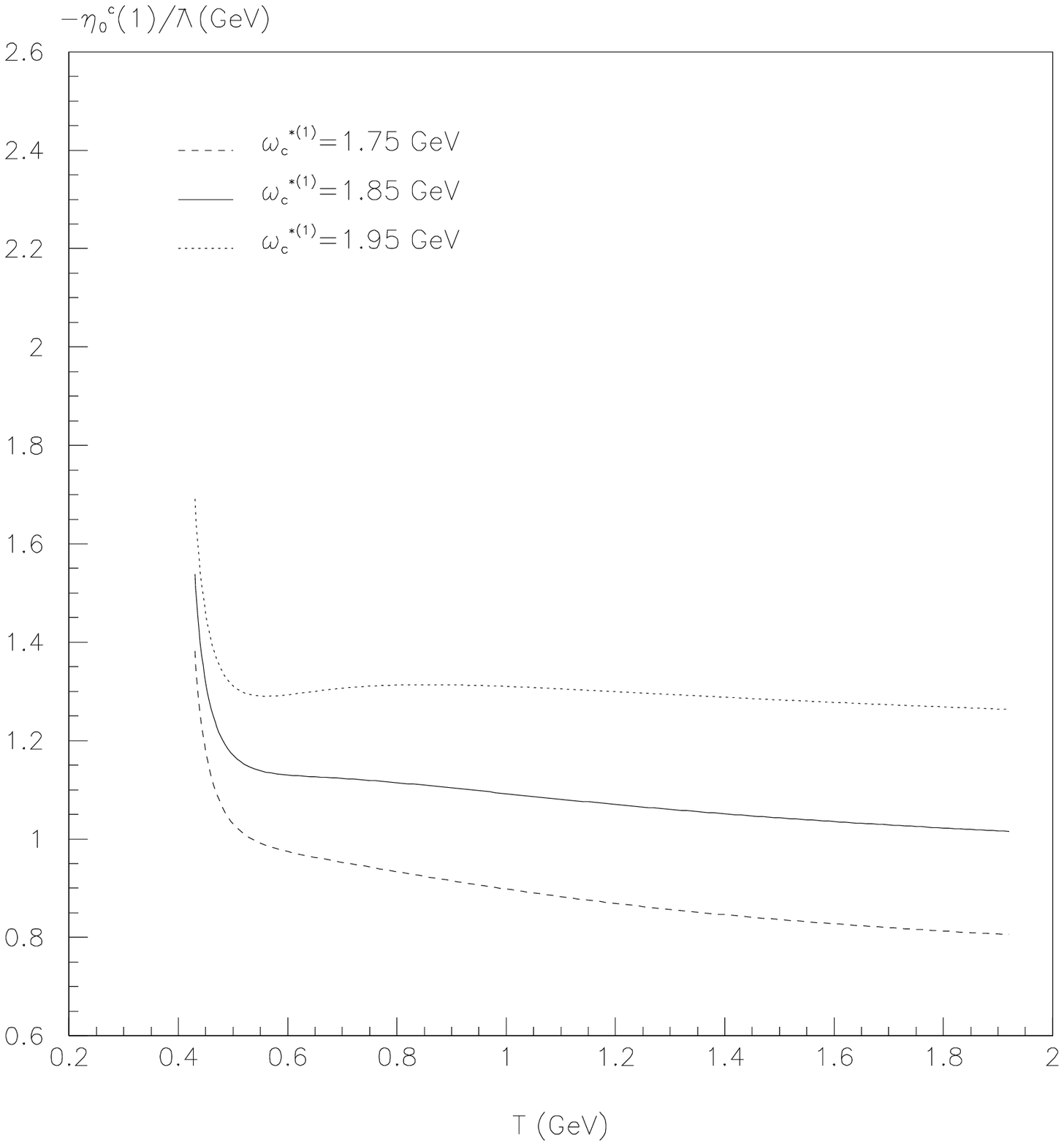}{Fig.2c}

\vspace{-1cm}{\centerline{Fig.2. Wave functions $\tau$, $\eta^b_0$ and $\eta^c_0$ at the 
zero recoil point $y=v\cdot v'=1.$}}

\vspace{1cm}

\PIC{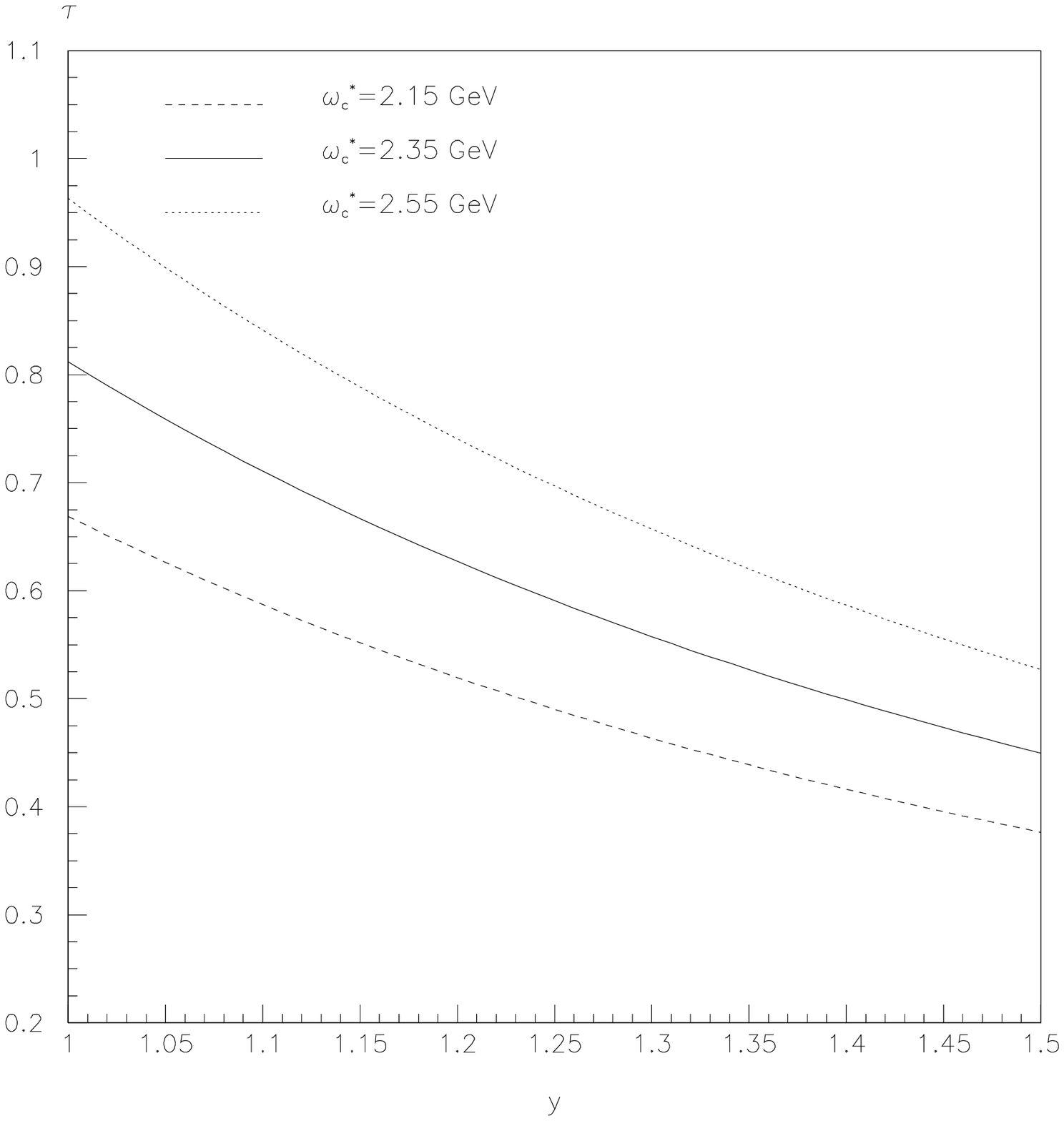}{Fig.3a}

\PIC{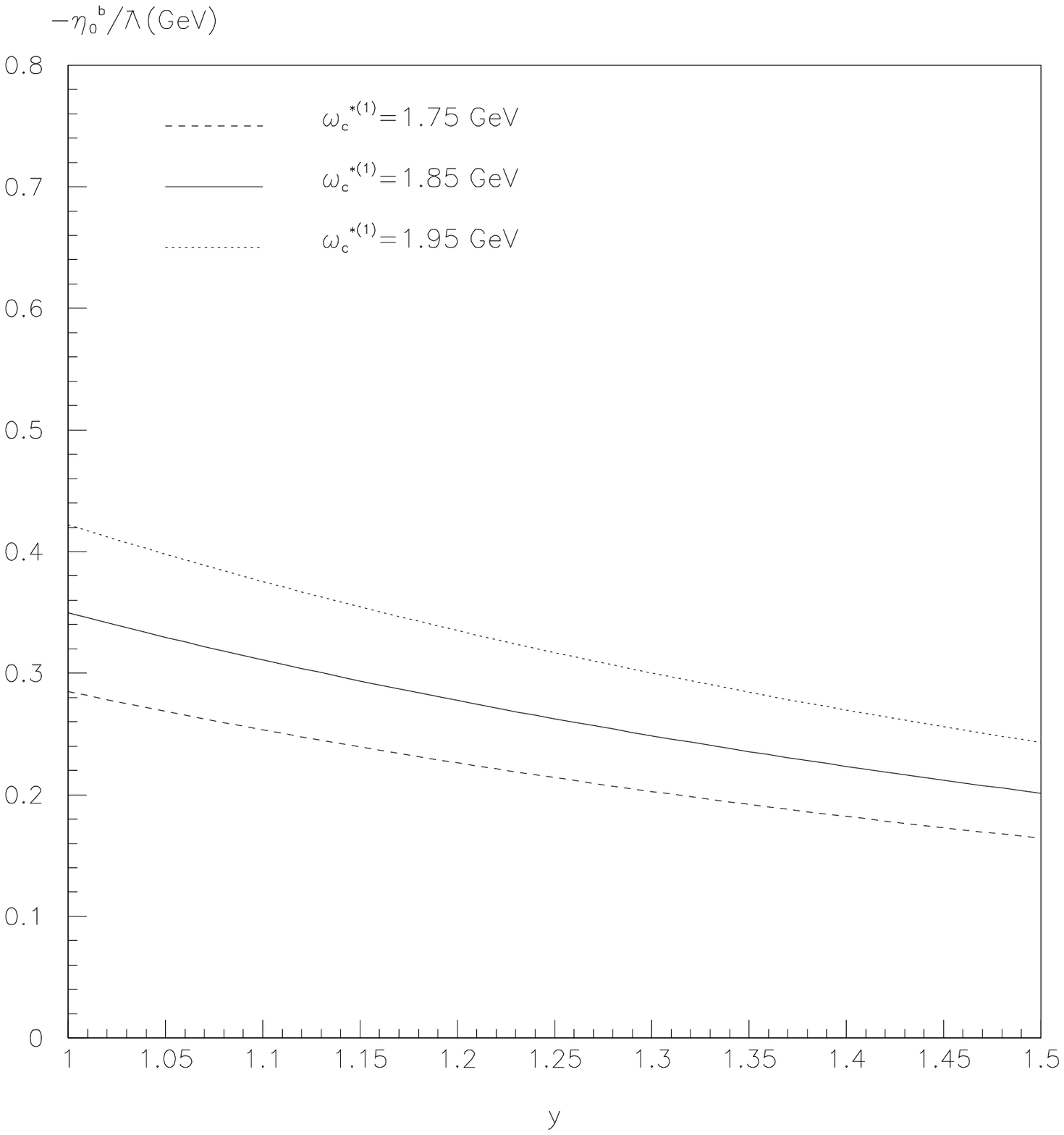}{Fig.3b}

\PIC{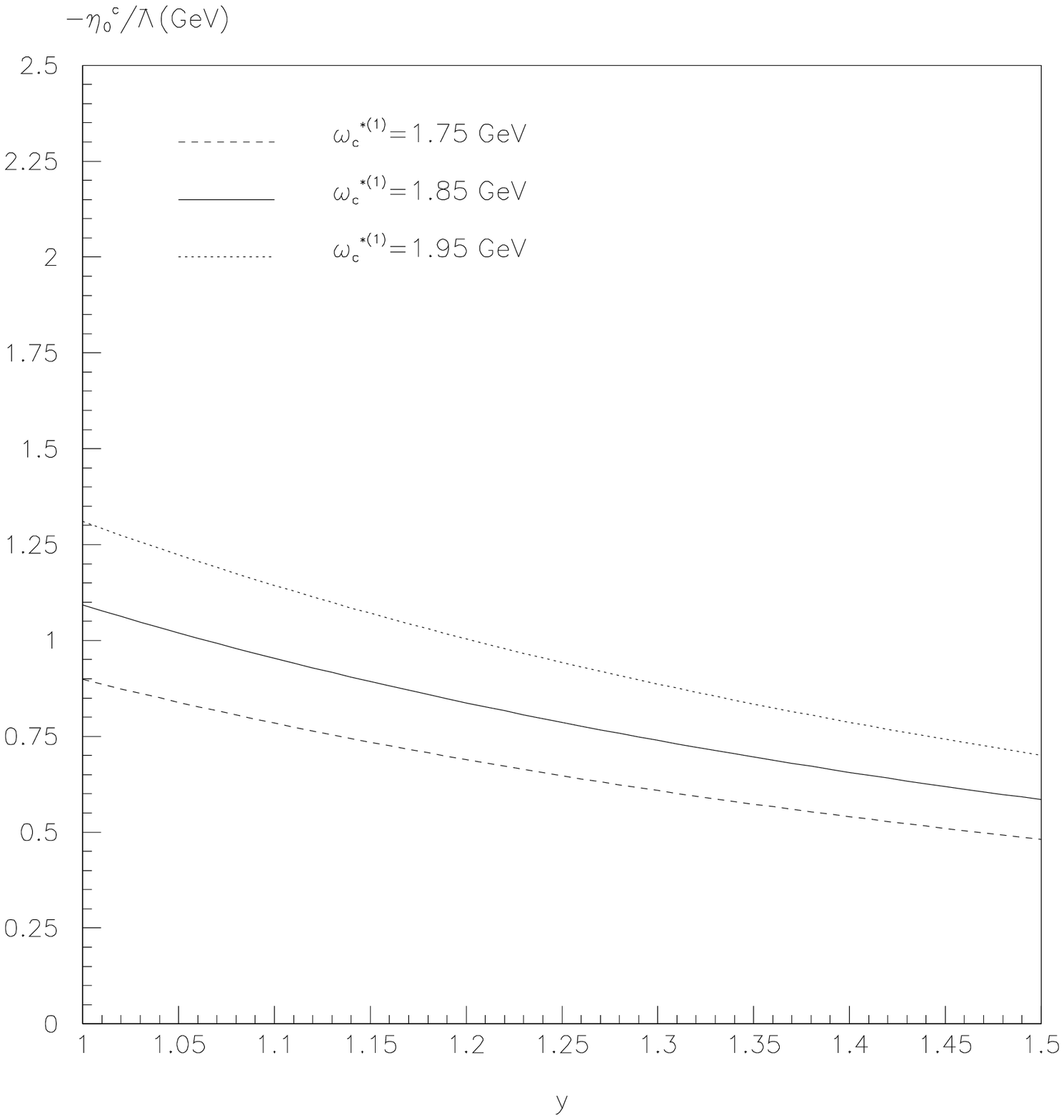}{Fig.3c}

\vspace{-1cm}{\centerline{Fig.3. Variations of the wave functions $\tau$, $\eta^b_0$ and $\eta^c_0$ 
with respect to $y$ at $T=1$GeV.}
}

\vspace{2cm}

\PIC{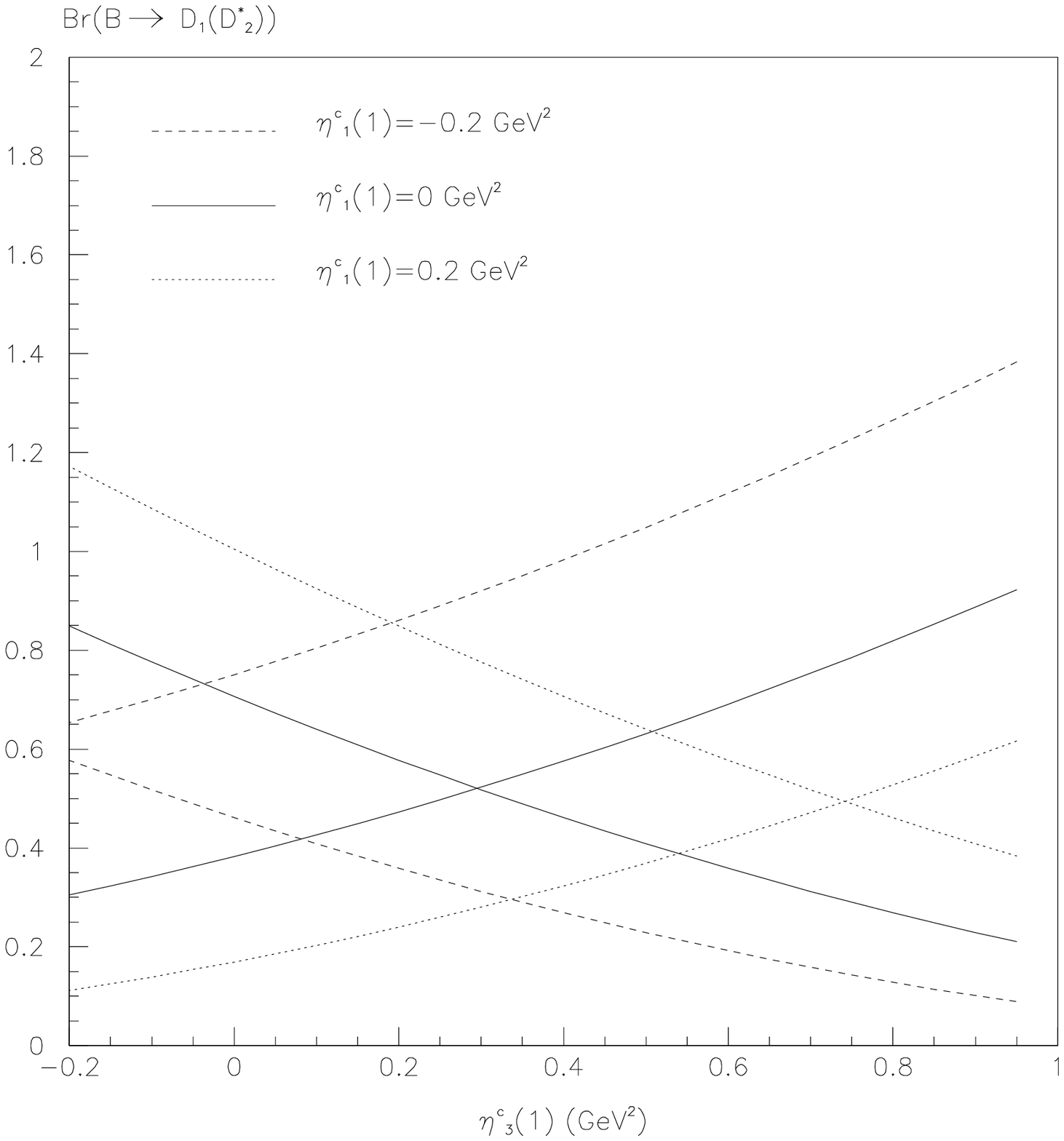}{Fig.4a}

\PIC{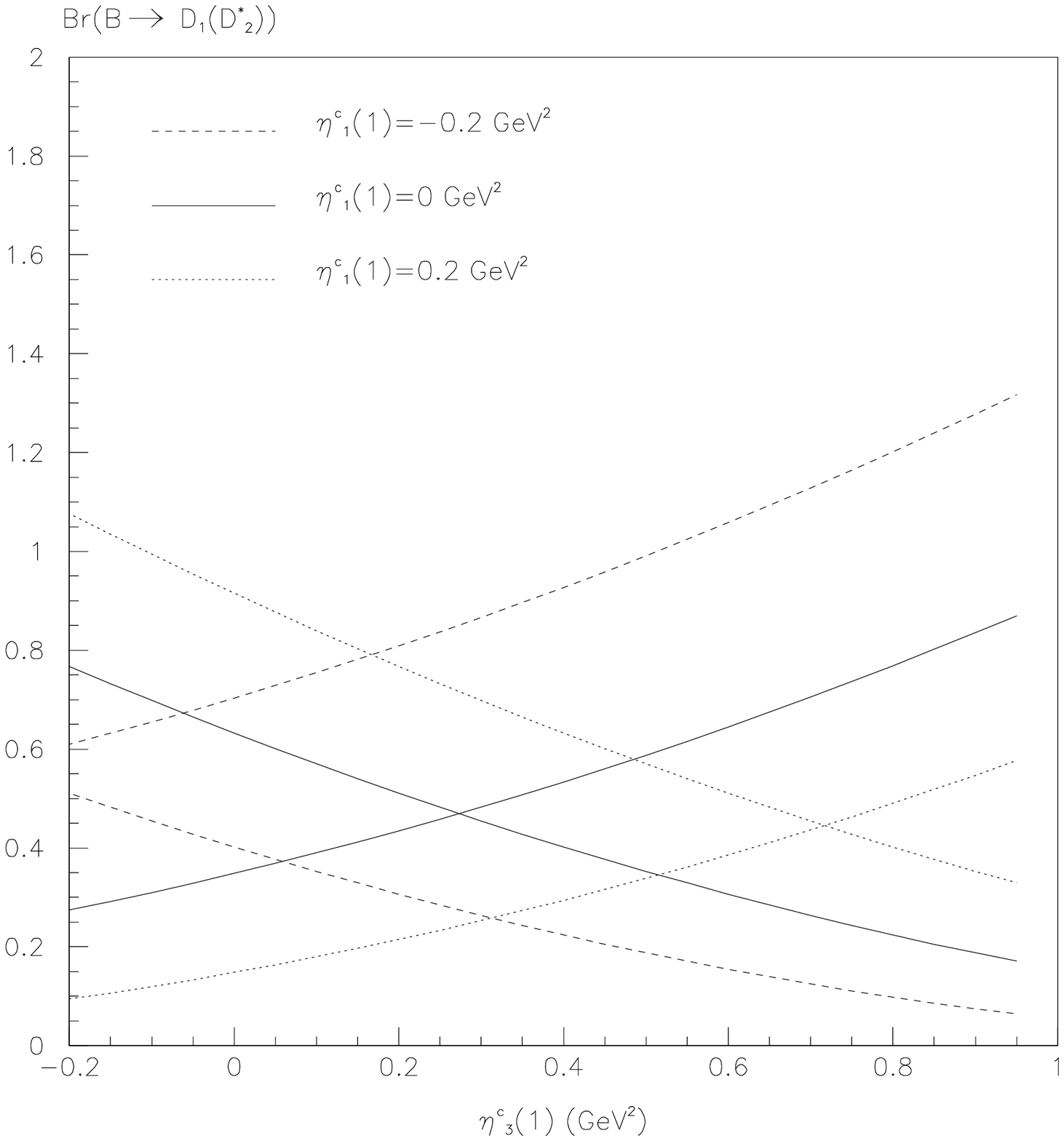}{Fig.4b}

\PIC{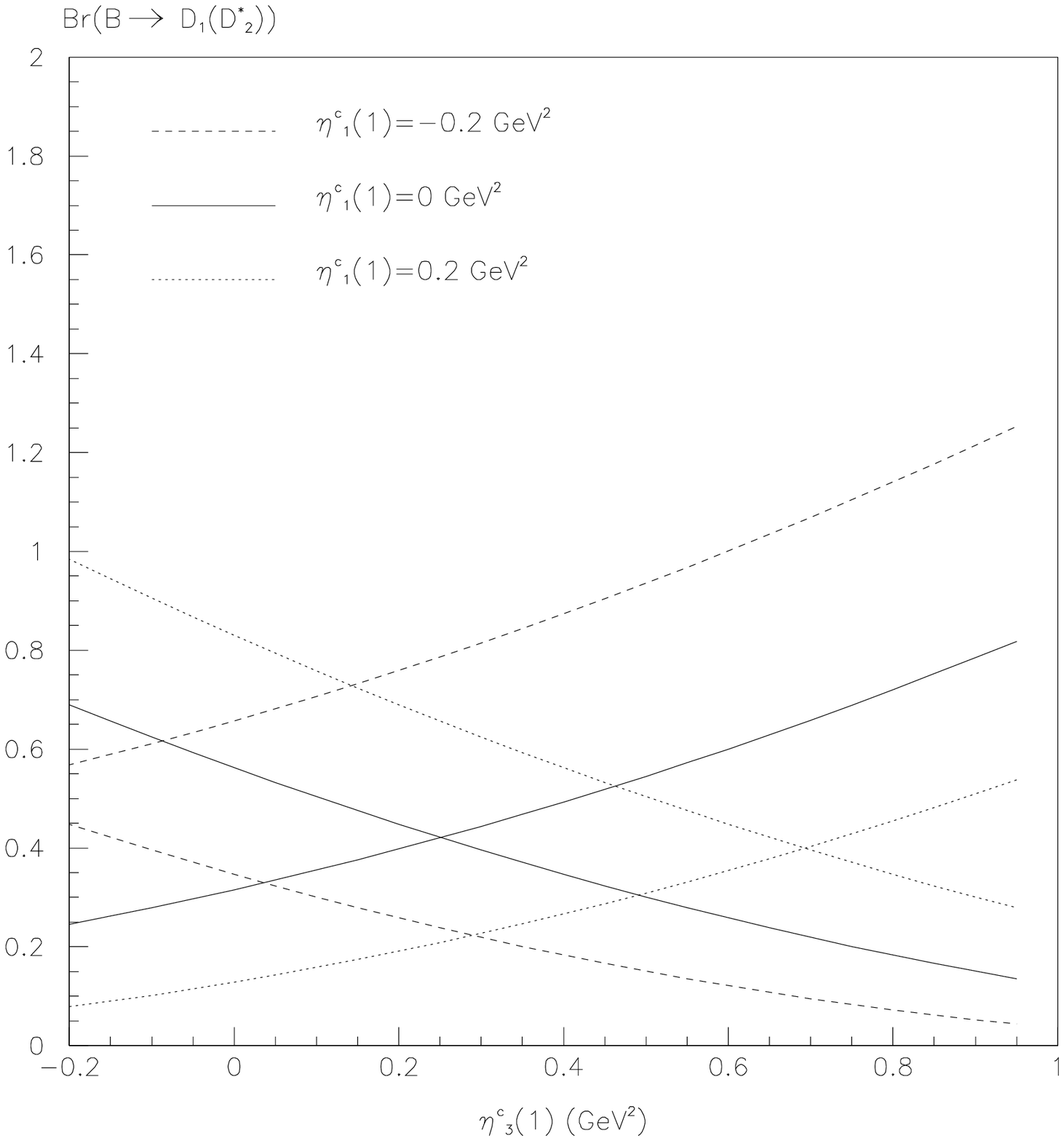}{Fig.4c}

\vspace{-2cm}{\center{Fig.4. Branching ratios of $B\to D_1(D^*_2)$ decays 
at different values of $\eta^Q_i$. a. $\eta^b(1)=-0.4\mbox{GeV}^2$; 
b. $\eta^b(1)=-0.6\mbox{GeV}^2$; c. $\eta^b(1)=-0.8\mbox{GeV}^2$. In these figures 
$\eta^c_2(1)=-0.6\mbox{GeV}^2$ is used, and we assumed that $\eta^Q_i$ and $\tilde\tau$ 
have the same monopole.}
}

\newpage

Table 1. Decay rates $\Gamma$ (in units of $|V_{cb}/0.038|^2\times 10^{-15}$ GeV) 
and branching ratios BR (in \%) for $B\to D_1(D^*_2)l\bar{\nu}$ decays in the infinitely 
heavy quark mass limit and taking account of first order $1/m_Q$ corrections. $R$ is a ratio
of branching ratios taking account of $1/m_Q$ corrections to branching ratios in
the infinitely heavy quark mass limit.
\begin{center}
\begin{tabular}{c|c|c|c}
\hline \hline
    &    &   $B\to D_1l\bar\nu$ & $B\to D^*_2l\bar\nu$   \\
\hline
$m_Q\to \infty$ & $\Gamma$       &  $1.87 \pm 0.53$  & $3.24\pm 0.94 $ \\
\cline{2-4}
 & Br  & $ 0.45 \pm 0.13 $ & $0.79\pm0.23$\\
\hline
with $\eta^Q_0 $ & $\Gamma$       &  $2.42\pm 0.65 $ & $4.05 \pm 1.09 $\\
\cline{2-4}
and $\kappa^{(')}_i $ & Br  & $0.59 \pm 0.16$ & $0.99\pm 0.26$ \\
\hline
\multicolumn{1}{c}{$R$} && $ 1.30 \pm 0.54 $     & $ 1.25 \pm 0.05$  \\
\hline
Experiment & Br(CLEO) \cite{cleo}   & $ 0.56\pm 0.13 \pm 0.08 \pm 0.04 $ & $  \langle 0.8$ \\
\cline{2-4}
  & Br(ALEPH) \cite{aleph} & $0.74 \pm 0.16 $  & $  \langle 0.2$ \\
\hline  \hline
\end{tabular}
\end{center}

\vspace{3cm}

Table 2. Branching ratios (in \%) at some values of $\eta^Q_i$. $\eta^c_2(1)=-0.6\mbox{GeV}^2$ and the monopole 
assumption are used in deriving the branching ratios in this table. 
\begin{center}
\begin{tabular}{|c|c|c|c|c|}
\hline \hline
\label{sev}
$\eta^b(1)(\mbox{GeV}^2)$ &$ \eta^c_1(1)(\mbox{GeV}^2) $&$ \eta^c_3(1)(\mbox{GeV}^2)  $&$ Br(B\to D_1l\bar{\nu}) $&$ Br(B\to D^*_2l\bar{\nu}) $\\
\hline
$-0.4   $&$ -0.2 $&$ -0.1 $&$ 0.70 $&$ 0.52 $\\
\cline{2-5}   
 &$ -0.2 $&$ 0.1 $&$ 0.80 $&$ 0.41 $\\
\cline{2-5}   
 &$ 0 $&$ 0.2 $&$ 0.47 $&$ 0.58 $\\
\cline{2-5}   
 &$ 0 $&$ 0.4 $&$ 0.58 $&$ 0.46 $\\
\cline{2-5}   
 &$ 0.2 $&$ 0.4 $&$ 0.32 $&$ 0.71 $\\ 
\cline{2-5}   
 &$ 0.2 $&$ 0.6 $&$ 0.42 $&$ 0.58 $\\ 
\hline
$-0.6 $&$ -0.2 $&$ -0.1 $&$ 0.66 $&$ 0.45 $\\
 \cline{2-5}   
  &$ -0.2 $&$ 0.1 $&$ 0.75 $&$ 0.35 $\\
 \cline{2-5}   
  &$ 0 $&$ 0.2 $&$ 0.43 $&$ 0.51 $\\
 \cline{2-5}   
  &$ 0 $&$ 0.4 $&$ 0.53 $&$ 0.40 $\\
 \cline{2-5}   
  &$ 0.2 $&$ 0.4 $&$ 0.29 $&$ 0.63 $\\
 \cline{2-5}   
  &$ 0.2 $&$ 0.6 $&$ 0.39 $&$ 0.51 $\\
\hline
$ -0.8 $&$ -0.2 $&$ -0.1 $&$ 0.61 $&$ 0.40 $\\
  \cline{2-5}   
  &$ -0.2 $&$ 0.1 $&$ 0.71 $&$ 0.30 $\\
  \cline{2-5}   
  &$ 0 $&$ 0.2 $&$ 0.40 $&$ 0.45 $\\
 \cline{2-5}   
  &$ 0 $&$ 0.4 $&$ 0.49 $&$ 0.35 $\\
 \cline{2-5}   
  &$ 0.2 $&$ 0.4 $&$ 0.27 $&$ 0.56 $\\
 \cline{2-5}   
  &$ 0.2 $&$ 0.6 $&$ 0.35 $&$ 0.45 $\\
\hline \hline
\end{tabular}
\end{center}

\vspace{3cm}

Table 3. Branching ratios (in \%) at some values of $\eta^Q_i$. $\eta^c_2(1)=0\mbox{GeV}^2$ and the monopole 
assumption are used in deriving the branching ratios in this table. 
\begin{center}
\begin{tabular}{|c|c|c|c|c|}
\hline \hline
\label{sev3}
$\eta^b(1)(\mbox{GeV}^2)$ &$ \eta^c_1(1)(\mbox{GeV}^2) $&$ \eta^c_3(1)(\mbox{GeV}^2)  $&$ Br(B\to D_1l\bar{\nu}) $&$ Br(B\to D^*_2l\bar{\nu}) $\\
\hline
$-0.4   $&$ -0.2 $&$ -0.1 $&$ 0.63 $&$ 0.61 $\\
\cline{2-5}   
 &$ -0.2 $&$ 0.1 $&$ 0.73 $&$ 0.49 $\\
\cline{2-5}   
 &$ 0 $&$ 0.2 $&$ 0.42 $&$ 0.67 $\\
\cline{2-5}   
 &$ 0 $&$ 0.4 $&$ 0.52 $&$ 0.55 $\\
\cline{2-5}   
 &$ 0.2 $&$ 0.4 $&$ 0.29 $&$ 0.81 $\\ 
\cline{2-5}   
 &$ 0.2 $&$ 0.6 $&$ 0.38 $&$ 0.67 $\\ 
\hline
$-0.6 $&$ -0.2 $&$ -0.1 $&$ 0.58 $&$ 0.54 $\\
 \cline{2-5}   
  &$ -0.2 $&$ 0.1 $&$ 0.68 $&$ 0.43 $\\
 \cline{2-5}   
  &$ 0 $&$ 0.2 $&$ 0.38 $&$ 0.60 $\\
 \cline{2-5}   
  &$ 0 $&$ 0.4 $&$ 0.48 $&$ 0.48 $\\
 \cline{2-5}   
  &$ 0.2 $&$ 0.4 $&$ 0.26 $&$ 0.73 $\\
 \cline{2-5}   
  &$ 0.2 $&$ 0.6 $&$ 0.35 $&$ 0.60 $\\
\hline
$ -0.8 $&$ -0.2 $&$ -0.1 $&$ 0.54 $&$ 0.48 $\\
  \cline{2-5}   
  &$ -0.2 $&$ 0.1 $&$ 0.63 $&$ 0.37 $\\
  \cline{2-5}   
  &$ 0 $&$ 0.2 $&$ 0.35 $&$ 0.53 $\\
 \cline{2-5}   
  &$ 0 $&$ 0.4 $&$ 0.44 $&$ 0.42 $\\
 \cline{2-5}   
  &$ 0.2 $&$ 0.4 $&$ 0.24 $&$ 0.66 $\\
 \cline{2-5}   
  &$ 0.2 $&$ 0.6 $&$ 0.32 $&$ 0.53 $\\
\hline \hline
\end{tabular}
\end{center}


\begin{references}
\bibitem{aklz} A. K. Leibovich, Z. Ligeti, I. W. Stewart and M. B. Wise, 
     Phys. Rev. Lett. {\bf 78}, 3995 (1997).
  \bibitem{akl} A. K. Leibovich, Z. Ligeti, I. W. Stewart and M. B. Wise, 
     Phys. Rev. D {\bf 57}, 308 (1998).
 \bibitem{dh} Y.B. Dai and M.Q. Huang, Phys. Rev. D {\bf 59}, 034018 (1999).
 \bibitem{dhl} M. Q. Huang, C. Z. Li and Y.B. Dai, Phys. Rev. D {\bf 61}, 054010 (2000).
   \bibitem{ylw} Y.L. Wu, Mod. Phys. Lett. {\bf A8}, 819 (1993).
   \bibitem{wwy1} W.Y. Wang, Y.L. Wu and Y.A. Yan,  Int. J. Mod. Phys. {\bf A15}, 1817 (2000); 
    hep-ph/9906529, 1999.
  \bibitem{ww} W.Y. Wang and Y.L. Wu, hep-ph/0006241.
  \bibitem{wwy2} Y.A. Yan, Y.L. Wu and W.Y. Wang, Int. J. Mod. Phys. {\bf A15}, 2735 (2000); 
    hep-ph/9907202, 1999.
  \bibitem{wy} Y.L. Wu and Y.A. Yan, to be published in Int. J. Mod. Phys. {\bf A}; 
    hep-ph/0002261, 2000.
  \bibitem{neu1076} M. Neubert, Phys. Rev. D {\bf 46}, 1076 (1992).
  \bibitem{ballnpb} P. Ball and V. M. Braun, Nucl. Phys. B {\bf 421}, 593 (1994). 
  \bibitem{aff} A.F. Falk, Nucl. Phys. B {\bf 378}, 79 (1992).
  \bibitem{dhh} Y.B. Dai, C.S. Huang, M.Q. Huang and C. Liu, 
    Phys. Lett. B {\bf 390}, 350 (1997); 
    Y.B. Dai, C.S. Huang and M.Q. Huang, Phys. Rev. D {\bf 55}, 5719 (1997); 
 \bibitem{neurep} M. Neubert, Phys. Rep. {\bf 245}, 259 (1994); 
 \bibitem{bs} B. Blok and M. Shifman, Phys, Rev. D {\bf 47}, 2949 (1993); 
 \bibitem{neubprd46} M. Neubert, Phys. Rev. D {\bf 46}, 3914 (1992).
 \bibitem{ebert} D. Ebert, R. N. Faustov and V. O. Galkin, Phys. Rev. D {\bf 61}, 014016 (2000); 
   hep-ph/9906415. 
 \bibitem{cleo} CLEO Collaboration, A. Anastassov et al., Phys. Rev. Lett. {\bf 80}, 
   4127 (1998).
 \bibitem{aleph} ALEPH Collaboration, D. Buskulic et al., Z. Phys. C {\bf 73}, 
   601 (1997).
   
\end{references}
\end{document}